\documentclass[twocolumn,amsmath,amssymb,superscriptaddress]{revtex4-1}

\usepackage{bm}
\usepackage[dvipdfmx]{graphicx}
\usepackage{braket} 
\usepackage{bbm} 
\usepackage{diagbox} 
\usepackage{colortbl} 
\usepackage{dcolumn}

\newcommand{\rr}{{\bm r}}
\newcommand{\RR}{{\bm R}}
\newcommand{\nn}{{\bm n}}
\renewcommand{\SS}{{\bm S}}

\renewcommand{\tt}{{\hat{\bm t}}}

\newcommand{\dn}{{\delta n}}
\newcommand{\bdn}{\delta {\bm n}}
\newcommand{\be}{{\bm\epsilon}}

\begin{document}

\title{Controllable inter-skyrmion attractions and resulting skyrmion-lattice structures in two-dimensional chiral magnets with in-plane anisotropy}

	\author{Mai Kameda}
	\affiliation{Institute for Materials Research, Tohoku University, Sendai 980-8577, Japan}
	\affiliation{Department of Applied Physics, Nagoya University, Nagoya 464-8603, Japan}
	
	\author{Rio Koyama}
	\affiliation{Department of Applied Physics, Nagoya University, Nagoya 464-8603, Japan}
	
	\author{Takuro Nakajima}
	\affiliation{Department of Applied Physics, Nagoya University, Nagoya 464-8603, Japan}

	\author{Yuki Kawaguchi}
	\affiliation{Department of Applied Physics, Nagoya University, Nagoya 464-8603, Japan}	

\date{\today}

\begin{abstract}
We study inter-skyrmion interactions and stable spin configurations in a 2D chiral magnet with in-plane anisotropies of a tilted magnetic field and the magneto-crystalline anisotropy on a (011) thin film.
We find that in both cases a small deformation of a skyrmion shape makes the inter-skyrmion interaction anisotropic, and that the skyrmions are weakly bounded along a certain direction due to an emergent attractive interaction.
Furthermore, when the magneto-crystalline anisotropy is comparable to the Zeeman energy, skyrmions embedded in a uniform magnetization are tightly bound by creating a magnetic domain between them.
The formation of the magnetic domain, and thus the strength of the inter-skyrmion interaction, can be controlled by the direction of an external magnetic field.
The anisotropic interaction also affects the skyrmion alignment in the skyrmion crystal (SkX) phase.
By employing the Monte Carlo simulation and the micromagnetic simulation,
we obtain an elongated triangular lattice structure in the SkX phase. In particular, 
in the presence of a strong magneto-crystalline anisotropy, magnetic domains appear in the background of the lattice structure, and bimerons aligned on the domain walls form an elongated triangular lattice.
We also find a parameter region that the SkX phase is stabilized due to the inter-skyrmion attraction.
\end{abstract}
\maketitle

\section{Introduction}
Magnetic skyrmions, nanometer-sized spin vortices, are appealing for their potential applications in magnetic memory and computing devices due to their topological stability~\cite{NagaosaTokura-nntech, Sai-Li-2021}.
A single skyrmion embedded in a uniform magnetization behaves as a particle characterized by a nonzero topological number ${\cal N} = \frac{1}{4\pi}\int dr^2\nn(\rr) \cdot(\partial_x\nn(\rr)\times\partial_y\nn(\rr))$, where $\nn(\rr)$ is a normalized spin vector at a position $\rr=(x,y)$. 
Skyrmions were originally proposed as elementary excitations by T. Skyrme in the field of nuclear physics ~\cite{Skyrme1962}, whereas skyrmions observed in chiral magnets, such as B20-type alloys $MX$ ($M =$ Mn, Fe, Co; $X =$ Si, Ge)~\cite{Muhlbauer-2009, Yu-2010} and $\beta$-Mn type Co-Zn-Mn alloys~\cite{Tokunaga-2015},
are stabilized by the Dzyaloshinskii-Moriya (DM) interaction~\cite{Dzyaloshinsky1958, Moriya1960}, 
and a crystal structure of magnetic skyrmions, called a skyrmion crystal (SkX), appears in thermal equilibrium~\cite{JungHoon-2017}.
Skyrmionic spin textures have been experimentally identified via the ac-susceptibility measurements \cite{Pfleiderer-1997}, neutron small angle scattering intensities for Fourier-space imaging \cite{Muhlbauer-2009}, Lorentz transmission electron microscopy for real-space imaging \cite{Yu-2010}, and the topological Hall effect \cite{Jiang-2016, Litzius-2016}.

Theoretical and experimental attempts have been made to expand skyrmion-hosting materials with the idea of utilizing a magnetic skyrmion as an information carrier.
Non-centrosymmetric magnets are the basic platform for realizing skyrmions, such as the above mentioned chiral magnets and the polar magnets GaV$_4$S$_8$ and GaV$_4$Se$_8$~\cite{Kezsmarki-2015, Bordacs-2017}, 
where the Bloch-type and N\'{e}el-type skyrmions are observed, respectively.
Here, the chiral magnet Cu$_2$OSeO$_3$~\cite{Seki-2012, Seki-2012-2,Adams-2012} and the polar magnets GaV$_4$S$_8$ and GaV$_4$Se$_8$ are multiferroic, and the ways of controlling skyrmion motions 
with electric field are discussed~\cite{Mochizuki-2015, Mochizuki-2015-2, Ruff-2015}.
Multilayer systems consisting of magnetic and heavy metal layers also realize strong DM interactions, such as iron mono-, bi-, and triple layers on an Ir substrate hosting atomic-scale skyrmions~\cite{Heinze-2011, Romming-2013, Romming-2015, Hanneken-2015, Hsu-2017}, and multilayer stacks of Pt/CoFeB/MgO, Pt/Co/Ta, and Pt/Co/MgO realizing skyrmions at room temperature~\cite{Woo-2016, Boulle-2016, Jiang-2016-2}.
More recently, the centrosymmetric magnets Gd$_2$PdSi$_3$~\cite{Kurumaji-2019} and GdRu$_2$Si$_2$~\cite{Khanh-2020} were found to host skyrmions: The former is due to a geometrically-frustrated triangular lattice~\cite{Okubo-2012} and the latter is attributed to four-spin interactions mediated by itinerant electrons.
The small-sized ($\sim$ 2 nm in diameter) skyrmions observed in these materials draw attention not only for the novel mechanism of stabilizing skyrmions but also for possible applications to high-density integration of magnetic storage.
We also note that anti-skyrmions with charge $\mathcal{N} = -1$~\cite{Nayak-2017,Peng-2020} and merons with charge $\mathcal{N} = 1/2$~\cite{Yu-2018,Nagase-2020}, in addition to the Bloch and N\'{e}el skyrmions with charge $\mathcal{N}=1$, have been demonstrated.

Focusing on chiral magnets, 
FeGe and Co-Zn-Mn alloys host stable or meta-stable skyrmions in the wide temperature range including room temperature and the wide magnetic field range up to $\sim$ 0.5~T~\cite{Tokunaga-2015, Zhao-2016, Karube-2016, Yu-2018, Karube-2018, Nagase-2019, Karube-2020}.
Zero-field robust skyrmions were also observed in FeGe~\cite{Karube-2017}.
In a bulk chiral magnet, the SkX phase appears only in a small region around the Curie temperature in the magnetic field--temperature phase diagram~\cite{Muhlbauer-2009}, whereas the region of the SkX phase is greatly enhanced down to 0~K in thin films~\cite{Yu-2010, Yu-2011, Tonomura-2012, Seki-2012, Leonov-2016}.

To improve device controllability, it would be crucial to manipulate inter-skyrmion interactions.
Here, we consider interactions between skyrmions embedded in a uniform background magnetization.
In a two-dimensional (2D) chiral magnet under a perpendicular magnetic field, the inter-skyrmion interaction is always repulsive and decays exponentially at a large distance~\cite{Piette1995, Shi-Zeng-2013}.
By considering three-dimensional (3D) magnetic structures in a bulk and a thin film, the attractive interactions between skyrmions are theoretically explained and indeed have been experimentally confirmed~\cite{Leonov-2016-2, Loudon-2018, Du-2018}.
The attractive interaction due to the softening of the magnetization near the transition temperature is also discussed~\cite{Wilhelm-2011}.
Besides chiral magnets, there are a few other mechanisms to introduce inter-skyrmion attractions:
Frustrated exchange interactions are shown to induce oscillation between repulsion and attraction \cite{Rozsa-2016, Lin-2016};
In a polar magnet with easy-plane anisotropy, skyrmions in a tilted ferromagnetic (FM) state undergoes anisotropic interactions and are bounded in a certain direction~\cite{Leonov-2017};
Biskyrmions, tightly bound pairs of skyrmions, observed in centrosymmetric magnetic films are attributed to the combined effect of the dipole-dipole interaction and the easy-axis anisotropy~\cite{Yu-2014, Wang-2016, Gobel-2019, Capic-2019}.
The interactions between skyrmions with higher topological numbers are also discussed in Refs.~\cite{Foster-2019, Capic-2020}.

In this paper, we theoretically investigate the 2D chiral magnet with in-plane anisotropy and show that 
there are two mechanisms to induce inter-skyrmion interactions, a distortion of skyrmion shape and the formation of a magnetic domain between skyrmions.
We analytically describe the inter-skyrmion interaction using a single-skyrmion solution, explaining the relation between skyrmion shape and interaction.
Based on the analytical expression, 
we consider two anisotropic effects that deform the skyrmion shape, (i) an in-plane magnetic field and (ii) the magneto-crystalline anisotropy, and numerically demonstrate that the inter-skyrmion attractions indeed appear.
In general, the magneto-crystalline anisotropy depends on the crystal plane direction to the film~\cite{Tokunaga-2015, Yu-2018, Nagase-2019, Nagase-2020}. 
We consider a (011) film to break the symmetry in spin space to create distorted skyrmions.
We also find that when the magneto-crystalline anisotropy is comparable to the Zeeman energy, the background magnetization is tilted from the perpendicular direction, and the skyrmions are tightly bound by forming a magnetic domain between them.
In particular, under the coexistence of the in-plane magnetic field and the magneto-crystalline anisotropy on the (011) film, the strength of the inter-skyrmion interaction is tunable in a wide range.
Such an external controllability of inter-skyrmion interactions proposed here may pave the way for further application of skyrmions.

We further investigate the SkX configurations 
and find unconventional states associated with the attractive couplings:
 a bimeron lattice formed on a background stripe domain pattern in the ground state and a one-dimensional (1D) skyrmion chain as an excitation in the FM phase.
Domain wall skyrmions and bimerons are already discussed and observed in the previous works~\cite{Cheng-2019, Xu-2020, Nagase-2020}. We here survey the optimal lattice structures in detail as a function of the external magnetic field and the strength of the magneto-crystalline anisotropy.
Notably, there is a magnetic field region where the lattice structure is sustained by the attractive interaction between the skyrmions or bimerons. 
In other words, the attractive inter-skyrmion interaction enhances the upper critical field for the SkX phase.

The paper is organized as follows. In Sec.~\ref{sec:analytical}, we introduce a continuum model of a 2D chiral magnet and analytically describe the inter-skyrmion interaction in terms of a single-skyrmion configuration. The detailed derivation is given in Appendices~\ref{sec:app_stereographic} and \ref{sec:app_interaction}.
We then explain that a deformation of skyrmions can induce an attractive coupling between them. In Sec.~\ref{sec:lattice_model}, a lattice model and a method of our micromagnetic simulation are described. 
In Sec.~\ref{sec:numerical_int_tiltB}, we discuss the inter-skyrmion interaction under a tilted magnetic field. By comparing the numerically obtained interaction and the approximate one derived in Sec.~\ref{sec:analytical}, we show that attractive interactions are indeed induced by a distortion of the skyrmion shape.
In Sec.~\ref{sec:numerical_int_aniso}, we discuss the inter-skyrmion interaction in the presence of the magneto-crystalline anisotropy. 
With weak anisotropy, we see small attractive inter-skyrmion interactions due to the skyrmion deformation, as in the case of Sec.~\ref{sec:numerical_int_tiltB}.
When the anisotropy becomes comparable to the Zeeman field, the stable FM state (uniform configuration) has a magnetization tilted from the Zeeman field, and a multiple magnetic domains are stabilized.
In this case, the inter-skyrmion attraction becomes considerably large by creating a magnetic domain between the skyrmions. 
In Sec.~\ref{sec:numerical_phase}, we investigate the ground-state phase diagram in the presence of the magneto-crystalline anisotropy. Some interesting skyrmion structures due to the inter-skyrmion attraction are discussed, such as the attraction-stabilized SkX phase and a 1D skyrmion chain in the FM phase.
In Sec.~\ref{sec:discussion}, we discuss several complemental issues, including the inter-skyrmion interactions on a (001) thin film and the combined effect of the in-plane magnetic field and the magneto-crystalline anisotropy.
Finally, we summarize the paper in Sec.~\ref{sec:conclusion}.

\section{Inter-skyrmion interaction: analytic approach}
\label{sec:analytical}

\subsection{General expression for the inter-skyrmion interaction}
We start from a continuum model for a thin film of a chiral magnet,
whose energy functional is given by
\begin{align}
F[\nn] &=\int \frac{d^2r}{a^2} f[\nn(\rr),\bm\nabla\nn(\rr)]\label{eq:energy_functional}\\
f[\nn(\rr),\bm\nabla\nn(\rr)] &= \frac{Ja^2}{2}[(\partial_x\nn)^2+(\partial_y\nn)^2]\nonumber\\
&\ \ +Da\nn\cdot(\bm \nabla\times \nn) +U_\textrm{c}(\nn,\bm\nabla\nn). \label{eq:f_CM}
\end{align}
Here, we choose the coordinate axes so that the film lies on the $x$-$y$ plane, $f$ is the energy per spin, $\nn(\rr)$ is a three-dimensional unit vector describing the direction of the magnetization, 
$J$ and $D$ are the strengths of the spin-exchange interaction and the DM interaction, respectively, $a$ is the lattice constant of the original lattice model (see next section), and $U_\textrm{c}(\nn,\bm\nabla\nn)$ is a function of $\nn$ and $\bm\nabla \nn=(\partial_x\nn,\partial_y\nn)$ that determines the anisotropy in the spin space.
We assume that the system has a uniform stationary solution $\nn(\rr)=\tt$. 
For example, when a magnetic field $B$ is applied in the direction of $\tt$,  the anisotropy potential $U_\textrm{c}(\nn,\bm\nabla\nn)=-B\tt\cdot\nn$ stabilizes the uniform solution.
The uniform stationary solution can be stable or metastable, appearing at least in the vicinity of the phase boundary between the FM phase and the SkX phase.
Our interest is the interaction between isolated skyrmions in such a region.

\begin{figure}
\begin{center}
\includegraphics[width=\linewidth]{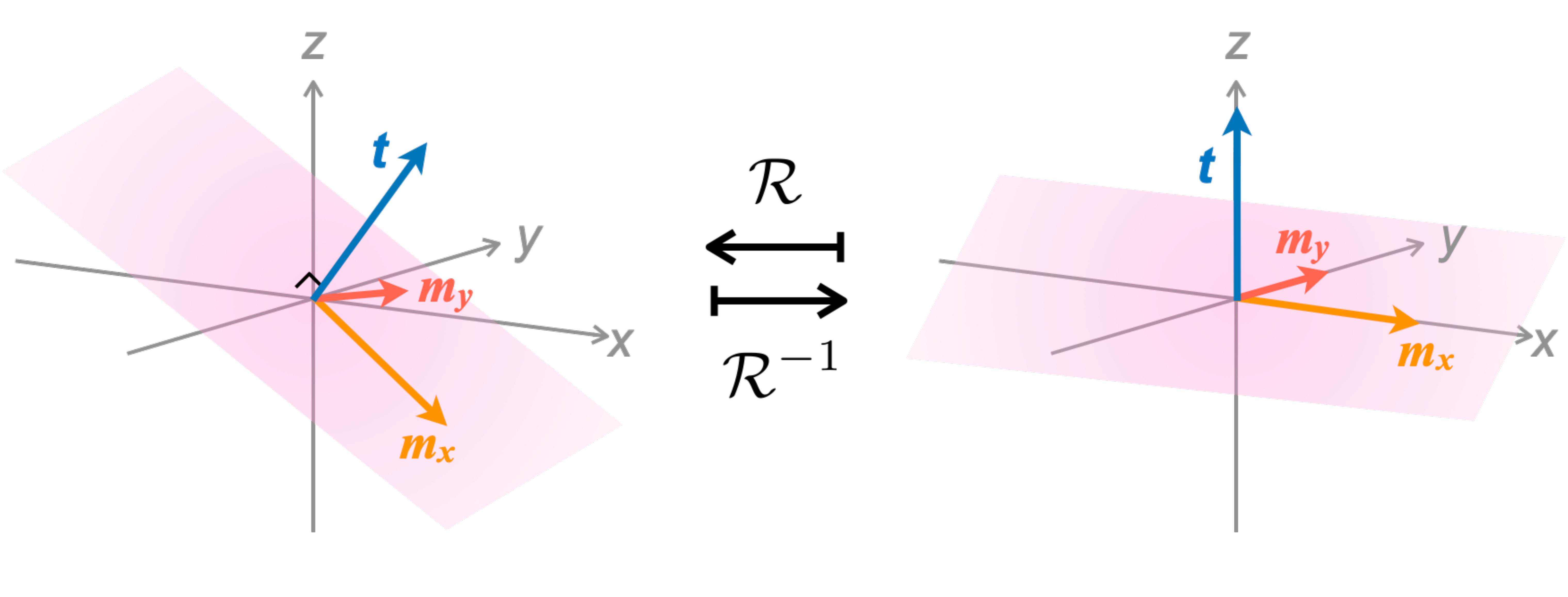}
\end{center}
\caption{
Rotation operation $\mathcal{R}$ and its inverse $\mathcal{R}^{-1}$ in the spin space.
Under the rotation $\mathcal{R}^{-1}$, $\tt$ is mapped to $\bm e_z$ and the plane perpendicular to $\tt$ is mapped to the $x$-$y$ plane.
The $x, y$, and $z$ components of the vector $\bm m=\mathcal{R}^{-1}(\bm n)$, defined after Eq.~\eqref{eq:Vint_J_2}, correspond to the  projection of $\bm n$ to the direction $\mathcal{R}\bm e_x$, $\mathcal{R}\bm e_y$ (which correspond to the directions of $m_x$ and $m_y$ in the left panel), and $\mathcal{R}\bm e_z=\tt$, respectively.
}
\label{fig:def_m1_m2}
\end{figure}

\begin{figure}
\begin{center}
\includegraphics[width=0.7\linewidth]{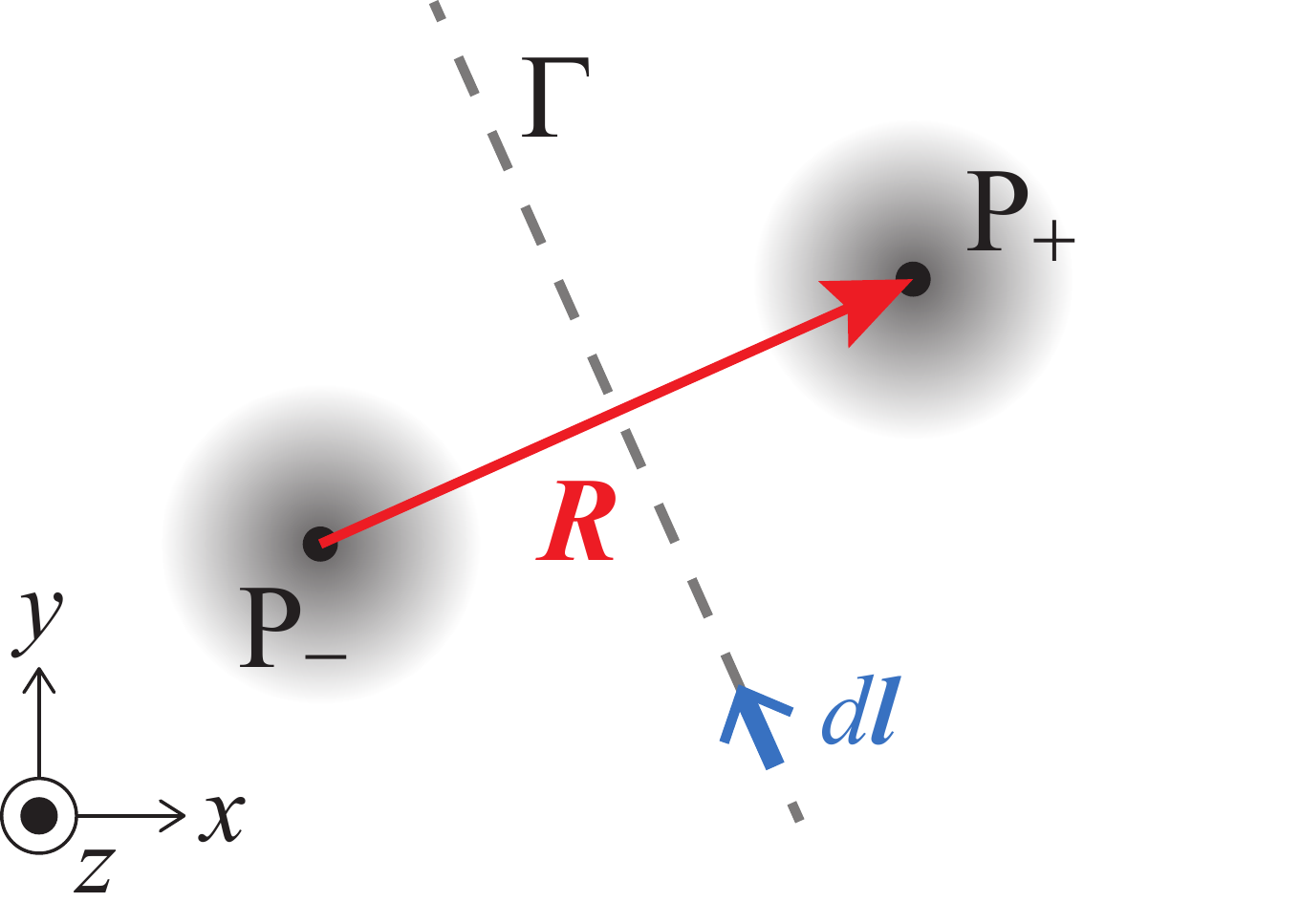}
\end{center}
\caption{
Schematic configuration for the calculation of the inter-skyrmion interaction, where two skyrmions are located at P$_+$ and P$_-$. 
The interaction between these skyrmions is approximated by the line integral along $\Gamma$ as given by Eq.~\eqref{eq:barV}.
}
\label{fig:config}
\end{figure}

We analytically evaluate the inter-skyrmion interaction at a distance.
Suppose that we have a stationary solution of a single-skyrmion state $\nn_{\rm 1sk}(\rr)$,
where a skyrmion at $\rr=\bm 0$ is embedded in a background uniform configuration,
i.e., $\nn_{\rm 1sk}(\bm 0)=-\tt$ and $\nn_{\rm 1sk}(\infty)=\tt$.
A state with a pair of skyrmions at points ${\rm P}_\pm:\rr=\pm\RR/2$ is obtained by summing up two vector fields
$\nn_\pm(\rr)=\nn_{\rm 1sk}(\rr\mp\RR/2)$ using the stereographic projection as follows~\cite{Piette1995}.
Let $\mathcal{R}:S^2 \mapsto S^2$ be a rotation operator about $\bm e_z\times \tt$ by an angle $\arccos(\bm e_z\cdot \tt)$, where $\bm e_{\alpha}$ $(\alpha=x,y,z)$ is the unit vector along the $\alpha$ axis.
As schematically shown in Fig.~\ref{fig:def_m1_m2}, the rotation $\cal{R}$ maps $\bm e_z$ to $\tt$, i.e., $\mathcal{R}{\bm e}_z=\tt$,
and the $x$-$y$ plane the plane orthogonal to $\tt$.
The stereographic projection,
$p:\mathbb{C}\cup\infty \mapsto S^2$, maps a complex number $u=u_1+iu_2$ to a three-dimensional unit vector as $p(u)=(2u_1,2u_2,1-|u|^2)/(1+|u|^2)$.
Then, the double-skyrmion state is described by $\nn_{\rm 2sk}=\mathcal{R} p[p^{-1} \mathcal{R}^{-1}(\nn_+)+p^{-1} \mathcal{R}^{-1}(\nn_-)]$. See Appendix~\ref{sec:app_stereographic} for more details.
The inter-skyrmion interaction potential is given by the energy difference between a double-skyrmion state and two single-skyrmion states with respect to the uniform configuration:
\begin{align}
V(\RR)=\int \frac{d^2r}{a^2} \left[f(\nn_{\rm 2sk})-f(\nn_+)-f(\nn_-)+f(\tt)\right].
\label{eq:Vint}
\end{align}
After some calculations (see Appendix~\ref{sec:app_interaction}), we find that $V(\RR)$ at a distance is approximated by
\begin{align}
{V}_\textrm{app}(\RR)=&\frac{1}{a^2}\int_\Gamma \epsilon_{ij} (A_{-+}-A_{+-})_i dl_j,
\label{eq:barV}\\
(A_{+-})_i=&\frac{\partial^2f(\tt)}{\partial n_\alpha\partial (\partial_i n_\beta)}\dn_{+,\alpha}\dn_{\rm -,\beta}\nonumber \\
& +\frac{\partial^2f(\tt)}{\partial (\partial_k n_\alpha) \partial (\partial_i n_\beta)}(\partial_k\dn_{+,\alpha})\dn_{-,\beta}
\label{eq:def_A}
\end{align}
where $\Gamma$ is the perpendicular bisector of the segment ${\rm P_+P_-}$,
$d\bm\ell$ is the line element of $\Gamma$ in the direction of ${\bm e}_z\times\RR$ (see Fig.~\ref{fig:config}),
$\epsilon_{ij}$ is the Levi-Civita symbol,
and summation over repeated indices is implied, where Roman (Greek) indices denote the components in the coordinate (spin) space and take the values $x$ and $y$ ($x, y$ and $z$).
Here, we define $\bdn$ as the projected vector of $\nn$ on the plane perpendicular to $\tt$, i.e., 
$\bdn\equiv \nn-(\nn\cdot \tt)\tt$.
In the derivation of Eq.~\eqref{eq:def_A}, we have assumed that $\bdn_\pm$ on the path $\Gamma$ is small and approximated $\nn_\pm$ as 
$\nn_\pm=\sqrt{1-|\bdn_\pm|^2}\tt+\bdn_\pm\simeq \tt+\bdn_\pm$.
The approximate potential, Eqs.~\eqref{eq:barV} and \eqref{eq:def_A}, can be applied to other continuum spin models as long as they have a uniform FM state and a localized skyrmion in it as stable solutions. 

\subsection{Anisotropy potential}
In this paper, we consider the Zeeman field and the magneto-crystalline anisotropy as the anisotropy potentials.
The contribution of the Zeeman field to $U_\textrm{c}$ is given by
\begin{align}
U_\textrm{c}^\textrm{(Ze)}(\nn,\bm\nabla\nn)=-\bm B_\textrm{ex}\cdot \nn(\rr),
\label{eq:Uc_Zeeman}
\end{align}
where $\bm B_\textrm{ex}$ is a uniform external magnetic field.

The lowest-order magneto-crystalline anisotropy potential on a 3D cubic lattice 
is written in the continuum model as~\cite{Bak-1980}
\begin{align}
    &U_\textrm{c}^\textrm{(mc,3D)}(\nn,\bm\nabla\nn)\nonumber\\
    &=\sum_{\nu=1,2,3}
    \left\{ A (\nn\cdot\hat{\bm p}_\nu)^4 - \frac{Ka^2}{2}[\partial_\nu (\nn\cdot\hat{\bm p}_\nu)]^2
    \right\},
\end{align}
where $A$ and $K$ are the strengths of the anisotropy, $\hat{\bm p}_{1,2,3}$ are the unit vectors pointing the three crystalline axes,
and $\partial_\nu$ denotes the derivative along $\hat{\bm p}_\nu$.
In a 2D film whose width along the $z$ axis is thin enough, $\partial_z$ appearing in $\partial_{\nu=1,2,3}$ is negligible.
For example, the magneto-crystalline anisotropy in a (011) thin film, which we discuss in the following, is described with $\hat{\bm p}_1={\bm e}_x$, $\hat{\bm p}_2=({\bm e}_y+{\bm e}_z)/\sqrt{2}$, and 
$\hat{\bm p}_3=(-{\bm e}_y+{\bm e}_z)/\sqrt{2}$.
The resulting anisotropy potential is given by
\begin{align}
    &U_\textrm{c}^\textrm{(mc,011)}(\nn,\bm\nabla\nn)\nonumber\\
    &=A 
    \left[n_x^4+ \frac{(n_y+n_z)^4}{4} + \frac{(-n_y+n_z)^4}{4}\right]\nonumber\\
    &- \frac{Ka^2}{4} \left[ 2(\partial_x n_x)^2+(\partial_y n_y)^2+(\partial_y n_z)^2 \right].
    \label{eq:Uc_mc}
\end{align}

\begin{figure}[htbp]
    \begin{tabular}{l}
          \includegraphics[width=0.85\linewidth]{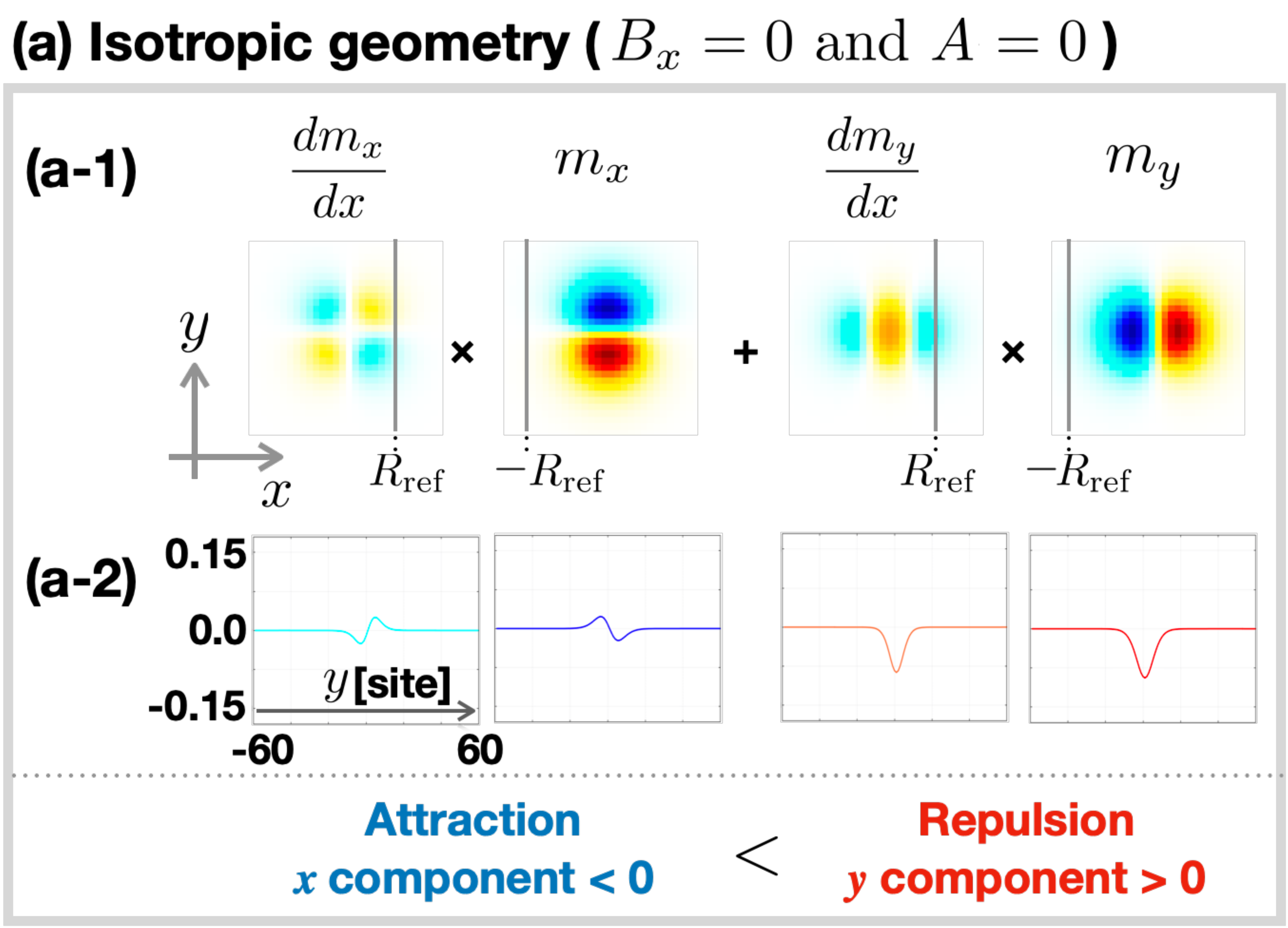}\\[1mm]
          \includegraphics[width=0.85\linewidth]{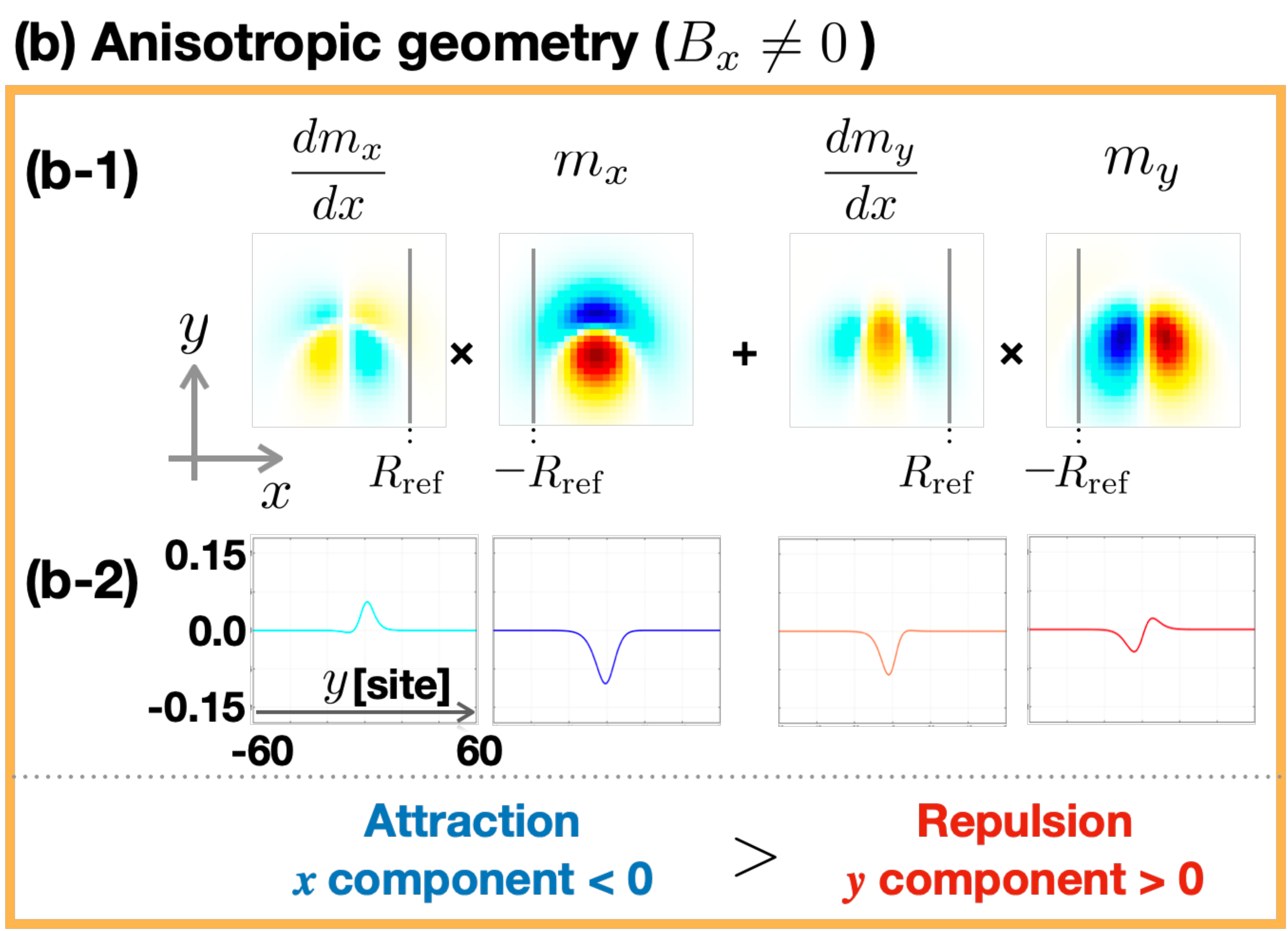}\\[1mm]
          \includegraphics[width=0.972\linewidth]{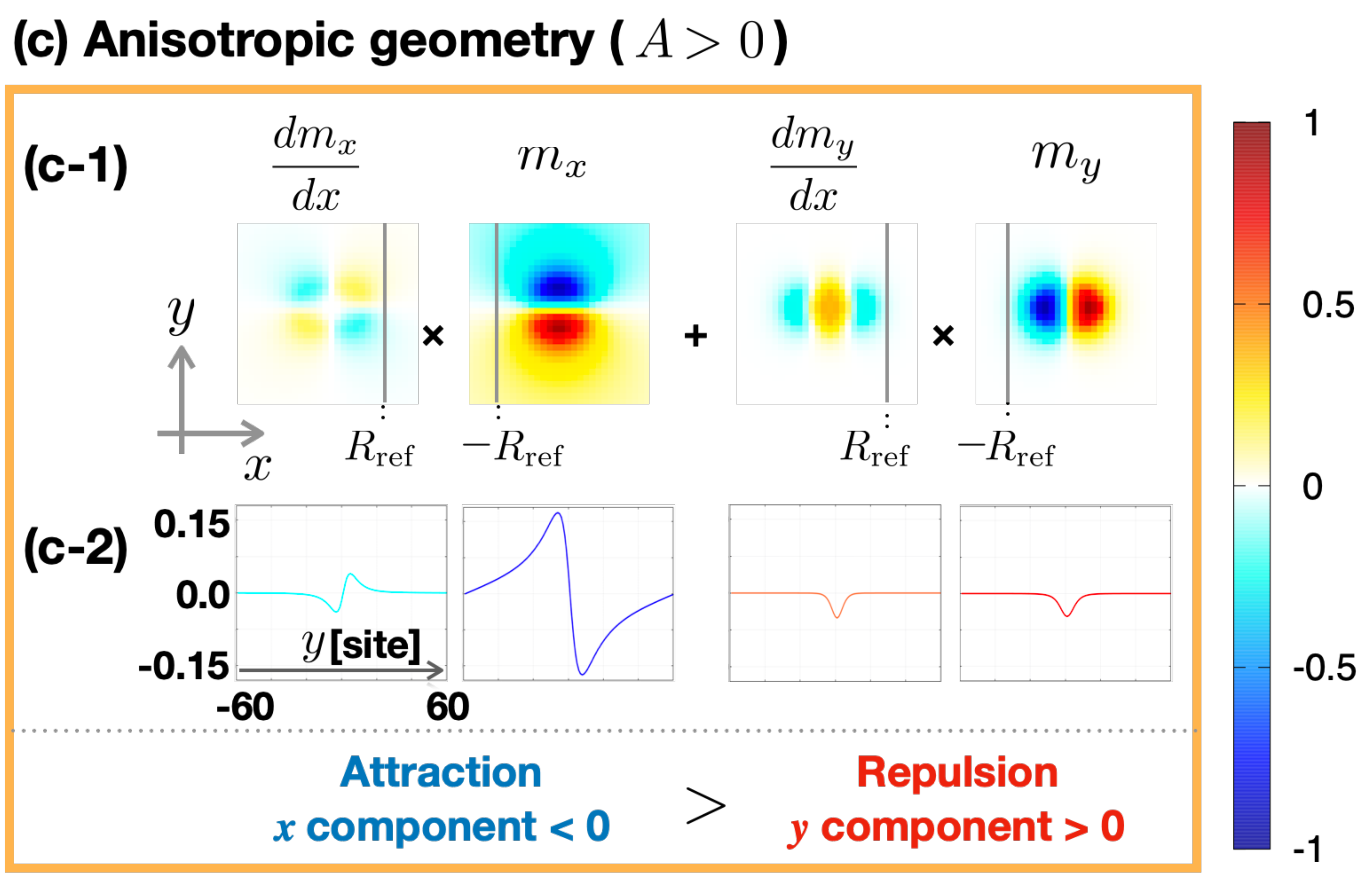}
    \end{tabular}
    \caption{
    Comparison of the contributions from the $x$ and $y$ components to the integral, Eq.~\eqref{eq:Vint_J_2}, 
    in (a) isotropic and (b), (c) anisotropic geometries. We numerically calculate the stationary single-skyrmion state $\bm n_\textrm{1sk}$ under the anisotropic potential $U_\textrm{c}=U_\textrm{c}^\textrm{(Ze)}+U_\textrm{c}^\textrm{(mc,011)}$, where $U_\textrm{c}^\textrm{(Ze)}$ and $U_\textrm{c}^\textrm{(mc,011)}$ are given in Eqs.~\eqref{eq:Uc_Zeeman} and \eqref{eq:Uc_mc}, respectively. Panels (a)-(c) are the results for (a) $\bm B_\textrm{ex}\parallel\bm e_z$ and $A=K=0$, (b) $\bm B_\textrm{ex}\parallel (\sin 30^\circ, 0,\cos 30^\circ)$ and $A=K=0$, and (c) $\bm B_\textrm{ex}\parallel\bm e_z$, $A\neq 0$, and $K=0$. (a-1)-(c-1) Color plots of $
   \partial_x m_x, m_x, \partial_x m_y$, and $m_y$ in the $x$-$y$ plane, where $\bm m\equiv \mathcal{R}^{-1}(\bm n_\textrm{1sk})$. The size of each panel is 30 sites $\times$ 30 sites. (a-2)--(c-2) $y$ dependence of $\partial_x m_x(R/2,y), m_x(-R/2,y), \partial_x m_y(R/2,y)$, and $m_y(-R/2,y)$ from left to right at $R/2=R_\mathrm{ref}$ = 10. In all cases of (a)--(c), the integrals of $\partial_x m_x(R/2,y) m_x(-R/2,y)$ and $\partial_x m_y(R/2,y) m_y(-R/2,y)$ for $y$ result in negative and positive values, respectively. The sum of them is positive for (a) and negative for (b) and (c), indicating repulsive and attractive inter-skyrmion interactions, respectively.
}
    \label{fig:Vr_Sk_shape}
\end{figure}

\subsection{Inter-skyrmion interaction in an isotropic geometry}
Using Eq.~\eqref{eq:f_CM} with the anisotropic potential $U_\textrm{c}=U_\textrm{c}^\textrm{Ze}+U_\textrm{c}^\textrm{(mc,011)}$, Eq.~\eqref{eq:def_A} reduces to
\begin{align}
&(A_{+-})_i\nonumber\\
&=Ja^2(\partial_i\bdn_+)\cdot\bdn_-\nonumber\\
&-Da (\bdn_+\times \bdn_-)_i\nonumber\\
&-Ka^2(\partial_x\delta n_{+,x})\delta n_{-,x}\delta_{i,x}\nonumber\\
&-\frac{Ka^2}{2}\left[(\partial_y\delta n_{+,y})\delta n_{-,y}+(\partial_y\delta n_{+,z})\delta n_{-,z}\right]\delta_{i,y}.
\label{eq:A_CM}
\end{align}
Below, we discuss how the each term contributes to the interaction.

\subsubsection{Circular symmetric case}
We first consider the circular symmetric case where the external magnetic field is applied in the $z$ direction, $\bm B_\textrm{ex}=B{\bm e}_z$, and there is no magneto-crystalline anisotropy, $A=K=0$.
The background uniform solution for this setup is obviously given by $\tt={\bm e_z}$.
It follows that the contribution from the $D$ term of Eq.~\eqref{eq:A_CM} vanishes
since $\bdn_+\times \bdn_-\parallel\tt={\bm e_z}$.
Thus, only the $J$ term contributes to the inter-skyrmion interaction at a distance:
\begin{align}
V_\textrm{app}(\RR)=2J\int_\Gamma \epsilon_{ij}&
\left(\partial_i\bdn_-\right)\cdot\bdn_+ dl_j,
\label{eq:Vint_J}
\end{align}
where we did partial integration using $\bdn_\pm(\infty)=0$.
The obtained inter-skyrmion interaction is the same as that in the baby Skyrme model, which includes forth order terms of the spatial derivative in the energy functional so as to stabilize skyrmion solutions~\cite{Piette1995}.
Moreover, in the symmetric case as described in the above,
Eq.~\eqref{eq:Vint_J} is evaluated in the same manner as Ref.~\cite{Piette1995}:
By using the asymptotic form of a single skyrmion at a distance $\bdn_{\rm 1sk}(r,\varphi)\sim K_1(\sqrt{B/Ja^2}r)(-\sin\varphi,\cos\varphi,0)$,
we obtain a repulsive inter-skyrmion interaction $V_\textrm{app}(\RR)\propto Ja^2K_0(\sqrt{B/J}|\RR|)$,
where $(r,\varphi)$ is the polar coordinates about the center of the skyrmion, 
and the $K_n(z)$ is the modified Bessel function of $n$th order that has the asymptotic behavior $K_n(z)\sim \sqrt{\pi/2z}e^{-z}$ at $z\to\infty$.

\subsubsection{Effect of skyrmion deformation}
\label{sec:deformation_effect}
Here, we note that the repulsive inter-skyrmion interaction in the above case is resulting from a subtle energy balance between the $x$ and $y$ components of the inner product in the integrand of Eq.~\eqref{eq:Vint_J}.
To clarify this point, we choose $\RR=R{\bm e}_x$ and rewrite Eq.~\eqref{eq:Vint_J} as
\begin{align}
&V_\textrm{app}(R{\bm e}_x)\nonumber\\
&=2J\int_{-\infty}^\infty \sum_{\alpha=x,y}\left[\partial_x m_\alpha(R/2,y)\right]m_\alpha(-R/2,y) dy,
\label{eq:Vint_J_2}
\end{align}
where $\bm m\equiv \mathcal{R}^{-1}(\nn_\textrm{1sk})$, and $m_x$ and $m_y$ correspond to the components of $\bm n$ projected onto the perpendicular plane to $\tt$ (see Fig.~\ref{fig:def_m1_m2}). 
We note that when $\tt\parallel{\bm e}_z$, $m_x$ and $m_y$ are respectively equivalent to $\delta n_x$ and $\delta n_y$.
In Fig.~\ref{fig:Vr_Sk_shape}, we plot the $x$ and $y$ components of the numerically obtained $\bm m$ and $\partial_x\bm m$ in the $x$-$y$ plane and those along $x=\pm R_\textrm{ref}$ for various $U_\textrm{c}(\nn,\bm\nabla\nn)$, where we choose $R_\textrm{ref}$ to be close to the skyrmion radius.
Figure~\ref{fig:Vr_Sk_shape}(a) shows the result for $U_\textrm{c}(\nn,\bm\nabla\nn)=-Bn_z$, where the details of the numerical calculation shall be given in the next section. From Fig.~\ref{fig:Vr_Sk_shape}(a),  one can see that the product of the $x$ ($y$) components has a negative (positive) contribution to Eq.~\eqref{eq:Vint_J_2}. The summation of these terms gives small positive value, indicating a repulsive interaction.
We find that this subtle balance can be easily violated when the skyrmion structure deforms either by tilting the external magnetic field [Fig.~\ref{fig:Vr_Sk_shape}(b)] or by introducing the magneto-crystalline anisotropy [Fig.~\ref{fig:Vr_Sk_shape}(c)]. 
In Figs.~\ref{fig:Vr_Sk_shape}(b) and \ref{fig:Vr_Sk_shape}(c), the contribution from the $x$ ($y$) components increases (decreases) and the resulting inter-skyrmion interaction becomes attractive.

\subsubsection{Effect of the $D$ term}
\label{sec:D_terms}
When the background configuration is tilted from the $z$ axis, $\tt\neq\bm e_z$, 
the $D$ term in Eq.~\eqref{eq:A_CM} also contributes to the interaction.
We describe the contribution of the $D$ term with respect to $\bm m=\mathcal{R}^{-1}(\bm n_\textrm{1sk})$.
Suppose that two skyrmions are located at relative position $\RR=R\bm e_x$ in a background magnetization $\tt=(\cos\chi\sin\phi, \sin\chi\sin\phi,\cos\phi)$.
Using $\delta\bm n=\mathcal{R}(m_x,m_y,0)$,
the contribution of the $D$ term to the interaction is given by
\begin{align}
    &\frac{2D}{a}\int_{-\infty}^\infty dy(\delta \bm n_+\times\delta\bm n_-)_x \nonumber\\ 
    &=\frac{2D}{a} \sin\phi \cos\chi 
    \int_{-\infty}^\infty dy\nonumber\\
    &\ \ \ \left[m_x\left(-\frac{R}{2},y\right)m_y\left(\frac{R}{2},y\right)-m_x\left(\frac{R}{2},y\right)m_y\left(-\frac{R}{2},y\right)\right].
\label{eq:contribution_from_D-term}
\end{align}
When the skyrmion configuration is given by a simple spin rotation of that for $U_\textrm{c}(\bm n,\bm\nabla\bm n)=-Bn_z$, the integral in Eq.~\eqref{eq:contribution_from_D-term} vanishes because of the symmetry: $m_x(-R/2,y)=m_x(R/2,y)$ and $m_y(-R/2,y)=-m_y(R/2,y)$ [see Fig.~\ref{fig:Vr_Sk_shape}(a)].
Hence, an additional deformation of the skyrmion configuration is required for a nonzero contribution of the $D$ term.
Roughly speaking, the contribution from the $D$ term, Eq.~\eqref{eq:contribution_from_D-term}, is smaller than Eq.~\eqref{eq:Vint_J_2} by a factor $\sin\phi\cos\chi$.
The detailed values of these integrals depend on how the skyrmion deforms under an anisotropic geometry.
We will numerically show in Secs.~\ref{sec:numerical_int_tiltB} and \ref{sec:numerical_int_aniso} that the contribution of the $D$ term is small at large $R$ but becomes comparable to that from the $J$ term for small $R$.

\subsubsection{Effect of the $K$ term}
\label{sec:K_terms}
For the case of $K\neq 0$, the $K$ term in Eq.~\eqref{eq:A_CM} also contributes to the interaction.
To see the effect of the $K$ term, we assume that the background magnetization points the $z$ direction, i.e., $\tt=\bm e_z$, and consider the interaction of skyrmions aliened along the $x$ axis. 
The approximate interaction in this case is given by
\begin{align}
    &V_\textrm{app}(R{\bm e}_x)\nonumber\\
    &=2(J-K)\int_{-\infty}^\infty \left[\partial_x m_x(R/2,y)\right]m_x(-R/2,y) dy\nonumber\\
    &+2J\int_{-\infty}^\infty \left[\partial_x m_y(R/2,y)\right]m_y(-R/2,y) dy.
\label{eq:Vint_K}
\end{align}
Thus, the $K$ term modifies the weight of the $x$ component in Eq.~\eqref{eq:Vint_J_2}.
It follows that if $K$ is negative and satisfies $K<-J(I_x+I_y)/|I_x|$,
where $I_{\alpha}=\int_{-\infty}^\infty \left[\partial_x m_\alpha(R/2,y)\right]m_\alpha(-R/2,y) dy$,
the interaction energy becomes negative even when the skyrmion configuration is not distorted.
However, when we evaluate the above condition for the configuration shown in Fig.~\ref{fig:Vr_Sk_shape}(a), we obtain $K/J<-1.6$.
Such a strong anisotropy, although which is not realistic, accompanies the deformation of skyrmions, modifying the inter-skyrmion interaction via the $J$ term.
On the other hand, for a small anisotropy, $|K|\ll J$, its effect is mainly in deforming the skyrmion configuration, and the contribution of the $K$ term in Eq.~\eqref{eq:A_CM} is negligible, leading to qualitatively the same effect as other anisotropy effects.
Thus, in the following calculations, we choose $K=0$ for the sake of simplicity and investigate two situations
(i) under a tilted magnetic field and (ii) under the onsite anisotropy $A$.

\section{Numerical Method}
\label{sec:lattice_model}
\subsection{Model Hamiltonian}
To numerically survey inter-skyrmion interactions and stable spin configurations, we use the classical spin Hamiltonian on a square lattice given by
\begin{align}
H=&
\nonumber
-J\sum_{\rr} \SS_{\rr}\cdot (\SS_{\rr + {\bm e}_x}+\SS_{\rr + {\bm e}_y})\\
\nonumber
&-D\sum_{\rr} (\SS_{\rr}\times \SS_{\rr + {\bm e}_x} \cdot {\bm e}_x 
+\SS_{\rr}\times \SS_{\rr + {\bm e}_y} \cdot {\bm e}_y)\\
&+\sum_{\rr} U(\SS_{\rr}),
\label{eq:aisoHamiltonian}
\end{align}
where $\SS_{\rr}$ is the normalized spin vector on a site $\rr\in\{an_x{\bm e}_x+an_y{\bm e}_y\,|\,n_x,n_y\in\mathbb{Z}\}$, $J$ and $D$ are the same as those in the continuum model, and $U(\SS_\rr)$ is the anisotropy potential corresponding to $U_\textrm{c}(\nn,\bm\nabla\nn)$. 
The Hamiltonian~\eqref{eq:aisoHamiltonian} is the discretized expression of Eq.~\eqref{eq:f_CM} obtained by replacing $\nn(\rr)$, $\partial_i\nn(\rr)$, and $\int d^2 r/a^2$ with $\SS_\rr$, $(\SS_{\rr+{\bm e}_i}-\SS_\rr)/a$, and $\sum_\rr$, respectively.
Thus, when we refer to $\bm n(\rr)$ and the values described in terms of $\bm n(\rr)$ in the following sections, we evaluate them using the above replacement.

As an anisotropy potential, we consider the Zeeman field and the magneto-crystalline anisotropy of a (011) thin film.
The reason for choosing the (011) film rather than (001) is because the breaking of the $C_4$ symmetry is crucial for the skyrmion deformation that induces an attractive inter-skyrmion interaction.
Although the 2D lattice structure on a (011) plane of a cubic lattice is not a square one, we use a square lattice model constructed by discretizing $U_\textrm{c}^\textrm{(mc,011)}$ given in Eq.~\eqref{eq:Uc_mc} on a square lattice.
Such a treatment is valid when the skyrmion size is much larger than the lattice constant. 
The resulting anisotropy potential, including the Zeeman field of Eq.~\eqref{eq:Uc_Zeeman}, is given by
\begin{align}
    U(\SS_\rr)=&-\bm B_\textrm{ex}\cdot \SS_\rr \nonumber \\
    &+ A\left[(S^x_\rr)^4+\frac{(S^y_\rr+S^z_\rr)^4}{4}+\frac{(-S^y_\rr+S^z_\rr)^4}{4}\right]\nonumber\\
    &+ K \left[S^x_{\rr}S^x_{\rr + {\bm e}_x}+\frac{1}{2}(S^y_{\rr}S^y_{\rr + {\bm e}_y}+S^z_{\rr}S^z_{\rr + {\bm e}_y})\right].
\end{align}

In the rest of the paper, we independently discuss changes of the inter-skyrmion interaction due to (i) an in-plane magnetic field and (ii) the onsite magneto-crystalline anisotropy (the $A$ term).
In both calculations, we choose $K=0$ as discussed in Sec.~\ref{sec:K_terms}.
In case (i), we apply the in-plane magnetic field along the $x$ axis and use the anisotropy potential given by
\begin{align}
U_\textrm{i}(\SS_\rr)=-B( S_\rr^z\cos\phi+S_\rr^x\sin\phi),
\label{eq:Ui}
\end{align}
where $\phi$ is the angle between the external magnetic field to the $z$ axis.
In case (ii), we apply an external magnetic field perpendicular to the film and use the anisotropy potential
\begin{align}
U_\textrm{ii}(\SS_\rr)=&-BS_\rr^z\nonumber\\
&+A\left[(S^x_\rr)^4+\frac{(S^y_\rr+S^z_\rr)^4}{4}+\frac{(-S^y_\rr+S^z_\rr)^4}{4}\right].
\label{eq:Uii}
\end{align}

\subsection{Micromagnetic simulation}
\subsubsection{Inter-skyrmion interaction}
To numerically calculate the inter-skyrmion interaction, 
we first obtain the energy of a stationary state with a single skyrmion, $E_\textrm{1sk}$, and that with two skyrmions at relative position $\RR$, $E_\textrm{2sk}(\RR)$, as well as the energy of the fully spin polarized state $\SS_\rr=\tt$, $E_\textrm{ferro}$.
The stationary states are obtained by solving the Landau--Lifshitz--Gilbert (LLG) equation at absolute zero:
\begin{align}
\frac{d\SS_\rr}{dt} &= -\SS_\rr\times{\bm B}_{\mathrm{eff}} + \alpha\SS_\rr\times\frac{d\SS_\rr}{dt},
\label{eq:LLGeq}
\end{align}
where $\bm B_\mathrm{eff}=-\delta H/\delta \SS_\rr$ with $H$ given by Eq.~\eqref{eq:aisoHamiltonian} is the effective magnetic field, and $\alpha$ is the damping constant.
The positions of skyrmions are fixed by introducing a strong single-site pinning field at the center of skyrmions.
The inter-skyrmion interaction potential is given by
\begin{align}
    V(\RR)=E_\textrm{2sk}(\RR)-2E_\textrm{1sk}+E_\textrm{ferro},
\end{align}
which corresponds to Eq.~\eqref{eq:Vint} in the continuum model.

\subsubsection{Stable skyrmion lattice structure}
\label{sec:method_lattice}
To find the ground state of the Hamiltonian Eq.~\eqref{eq:aisoHamiltonian}, 
we combine the exchange Monte Carlo (MC)~\cite{Hukushima-1996} and the Metropolis MC methods. 
We first employ the exchange MC and seek thermal-equilibrium spin states in the temperature range of $0.01J\le k_\textrm{B}T\le J/2$,
where we empirically use 30 replicas. 
We then find the lowest-energy state at $k_\textrm{B}T=0.01J$ during a few tens of thousands of MC steps after the system is thermalized.
Setting the lowest-energy state as an initial state, 
we perform the Metropolis MC at $T=0$ 
to find the energy-minimum state.

\subsubsection{Parameter setup}
In the following calculation, we take $J=1, D=0.5$, and $a=1$. 
Under a perpendicular magnetic field in the absence of the magneto-crystalline anisotropy, 
the SkX phase arises in the magnetic field region of $B_\textrm{cr1}\le B \le B_\textrm{cr2}$, where the critical magnetic fields are obtained as $B_\textrm{cr1}\simeq 0.23D^2/J$ and $B_\textrm{cr2}\simeq 0.78D^2/J$~\cite{Iwasaki-2013, kawaguchi-2016}.
The lattice constant of the triangular SkX is given by $4\pi Ja/(\sqrt{3}D)$~\cite{NagaosaTokura-nntech}, which corresponds to the twice of the skyrmion radius $R_\textrm{sk}$ of an isolated skyrmion in the vicinity of the phase boundary between the FM and SkX phases. 
In our parameter setup, the skyrmion radius is given by $R_\textrm{sk}=7.3a$,
which is sufficiently larger than $a$, supporting the validity of using the square lattice model for a (011) thin film.

\section{INTER-skyrmion INTERACTION: Under Tilted External Magnetic Field}
\label{sec:numerical_int_tiltB}
In this section, we use the anisotropy potential $U_\textrm{i}(\SS_\rr)$ defined in Eq.~\eqref{eq:Ui} and show how the inter-skyrmion interaction changes as $\phi$ increases.
The previous work~\cite{Shi-Zeng-2015} has investigated the similar situation and shown that a skyrmion has a non-circular configuration and the inter-skyrmion interaction becomes anisotropic. We confirm these results and additionally find that the interaction becomes attractive at larger distance than the region the authors of Ref.~\cite{Shi-Zeng-2015} have investigated.

\begin{figure}[htb]
\begin{center}
\includegraphics[width=0.6\linewidth]{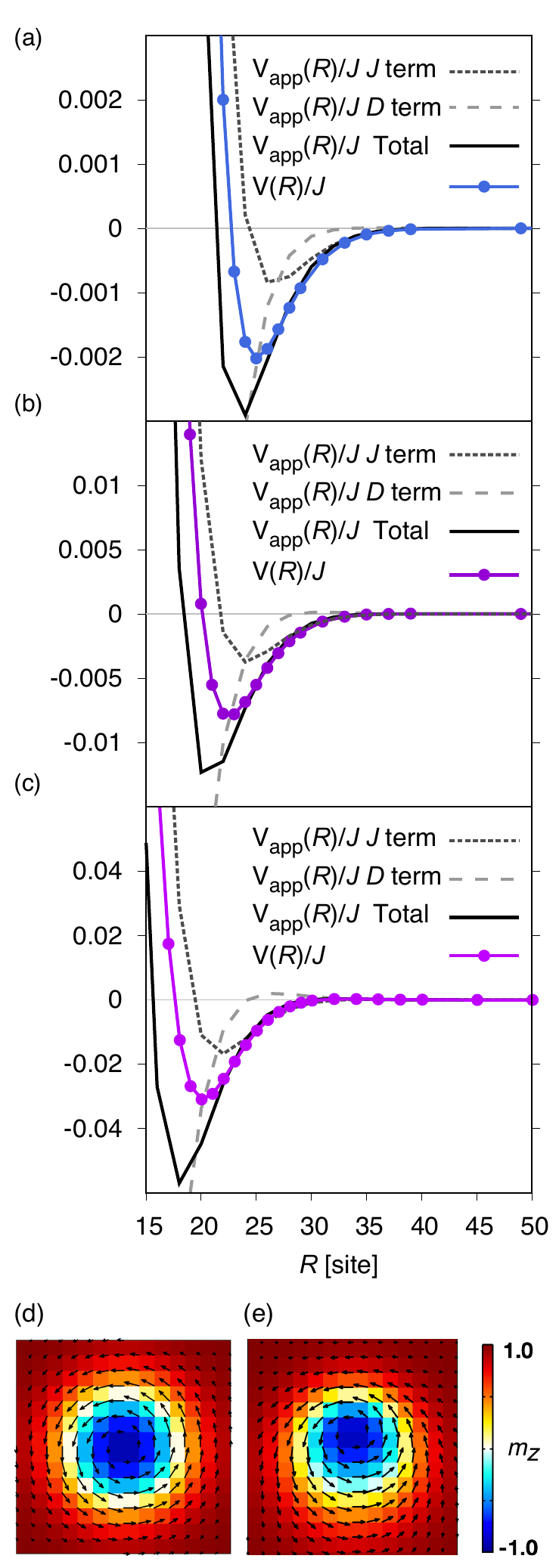}
\end{center}
\caption{(a)-(c)
Interaction potential $V(R)$ between two skyrmions aligned along the $x$ direction under a magnetic field of strength $BJ/D^2=0.73$ and tilting angle (a) $\phi=17^\circ$, (b) $\phi=22^\circ$, and (c) $\phi=30^\circ$. Shown are the numerically calculated interaction $V(R)$, the approximate one $V_\textrm{app}(R)$, and the contributions from the $J$ and $D$ terms in Eq.~\eqref{eq:A_CM}. (d),(e) Magnetization profile $\bm m=(\SS_\rr\cdot(\bm e_y\times \tt),S_{\bm \rr}^y,\SS_\rr\cdot \tt)$ of a single skyrmion configuration for (d) $\phi=0^\circ$ and (e) $\phi=30^\circ$, 
where the arrows represent the vector $\bm m$ projected on the $x$-$y$ plane and the color plot shows $m_z$.
Tilting of the external magnetic field does not merely cause the rotation in spin space but deforms the circular shape of the skyrmion, inducing attractive interaction.
}
\label{fig:tiltB_Vr}
\end{figure}

Figures~\ref{fig:tiltB_Vr}(a), (b), and (c) show the numerically obtained interaction potential $V(R)$ for skyrmions aligned along the $x$ direction at distance $R$ for $\phi=17^\circ, 22^\circ$, and $30^\circ$, respectively.
We also plot $V_\textrm{app}(R)$ defined in Eq.~\eqref{eq:barV}, as well as the contributions from the first and second terms of Eq.~\eqref{eq:A_CM} to $V_\textrm{app}(R)$, which are evaluated by using a numerically obtained single skyrmion configuration.
One can see that for all cases the interaction energy becomes negative for large $R$, which means that the inter-skyrmion interaction is attractive at a distance.  The magnitude of attractive interaction becomes larger for larger $\phi$, but the interaction energy is as small as a few percent of $J$.

The approximate interaction $V_\textrm{app}(R)$ well agrees with $V(R)$ for $R$ larger than that minimizes $V(R)$. The detailed comparison between them further reviles that the origin of the attraction at a large distance mainly comes from the $J$ term, as we discussed in Sec.~\ref{sec:D_terms}.
As $R$ becomes smaller, the contribution from the $D$ term becomes significant and comparable to that from the $J$ term at around the potential minimum.

The appearance of the attractive force, i.e., negative $V(R)$, can be understood from the deformation of a single skyrmion configuration.
In Fig. \ref{fig:Vr_Sk_shape}(b), we show $\bm m$ and $\partial_x\bm m$ for $\phi=30^\circ$. 
One can see that the distribution of $m_x$ ($m_y$) along the $x$ direction expands (contracts) compared with that for $\phi=0$ [Fig.~\ref{fig:Vr_Sk_shape}(a)]. Since the contributions to Eq.~\eqref{eq:Vint_J_2} from the $x$ ($y$) component is negative (positive), $V_\textrm{app}(R)$ for a fixed $R$ ($\gtrsim 2R_\textrm{sk}$) decreases as $\phi$ increases and eventually becomes negative.
One can also see that from Fig.~\ref{fig:Vr_Sk_shape}(b), Eq.~\eqref{eq:contribution_from_D-term} negatively contributes to the interaction potential.

We plot $\bm m=(\SS_\rr\cdot(\bm e_y\times \tt),S_{\bm \rr}^y,\SS_\rr\cdot \tt)$ of single-skyrmion configurations at $\phi=0^\circ$ and $30^\circ$ in Figs.~\ref{fig:tiltB_Vr}(d) and (e), respectively.
These figures indicate that the magnetization profile under a tilted magnetic field 
is not obtained by a simple rotation of the skyrmion configuration at $\phi=0$ in spin space but accompanies additional deformation, which leads to the interaction change.

\begin{figure}
\begin{center}
\includegraphics[width=\linewidth]{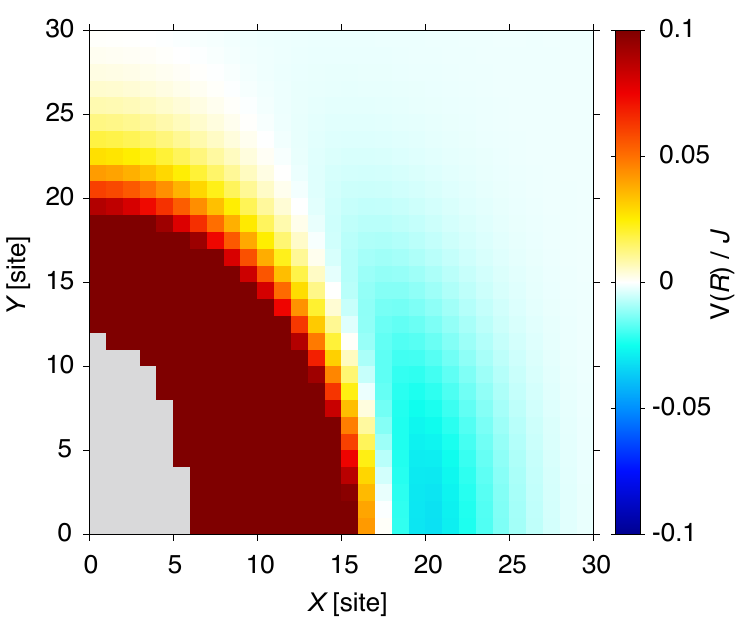}
\end{center}
\caption{
Interaction potential $V(\bm R)$ between two deformed skyrmions at relative position $\bm R=(X,Y)$ under a tilted magnetic field of strength $BJ/D^2=0.73$ and tilting angle $\phi=30^\circ$ in the $x$ direction.
}
\label{fig:aniso_Vr_tiltB}
\end{figure}

The interaction potential between deformed skyrmions depends on the relative direction as discussed in Ref.~\cite{Shi-Zeng-2015}.
Figure~\ref{fig:aniso_Vr_tiltB} shows the inter-skyrmion interaction potential as a function of the relative position $\bm R=(X,Y)$. The interaction potential has a minimum along the in-plane magnetic field, i.e., along the $x$ axis in the present case. On the other hand, the interaction energy along the $y$ axis increases as $\phi$ increases. The similar result is obtained in Ref.~\cite{Shi-Zeng-2015}.
However, Ref.~\cite{Shi-Zeng-2015} has investigated smaller region of $\bm R$ (up to 14 site in our parameter) and has not referred to the appearance of the attraction.

\section{INTER-skyrmion INTERACTION: Under Magnetic Anisotropy}
\label{sec:numerical_int_aniso}

In this section, we consider skyrmion deformation due to the magneto-crystalline anisotropy $U_\textrm{ii}(\SS_\rr)$ defined in Eq.~\eqref{eq:Uii}.
Note that when the crystalline anisotropy ($A$ term) dominates the Zeeman term, the magnetization direction of the uniform solution tilts from the $z$ axis. We first calculate the preferred direction $\tt$ of the uniform solution in Sec.~\ref{sec:Uii_direction}.
Then, we investigate the interaction between skyrmions embedded in the background magnetization $\tt=\bm e_z$ and $\tt\neq \bm e_z$ in Sec.~\ref{sec:Uii_interaction}.
Below, we consider only the case of $A>0$ because the qualitative behavior of the inter-skyrmion interaction are the same for $A>0$ and $A<0$ (see Sec.~\ref{sec:Uii_direction}).

\subsection{Preferred spin orientation due to magnetic anisotropy}
\label{sec:Uii_direction}
\begin{figure}
\begin{center}
\includegraphics[width=\linewidth]{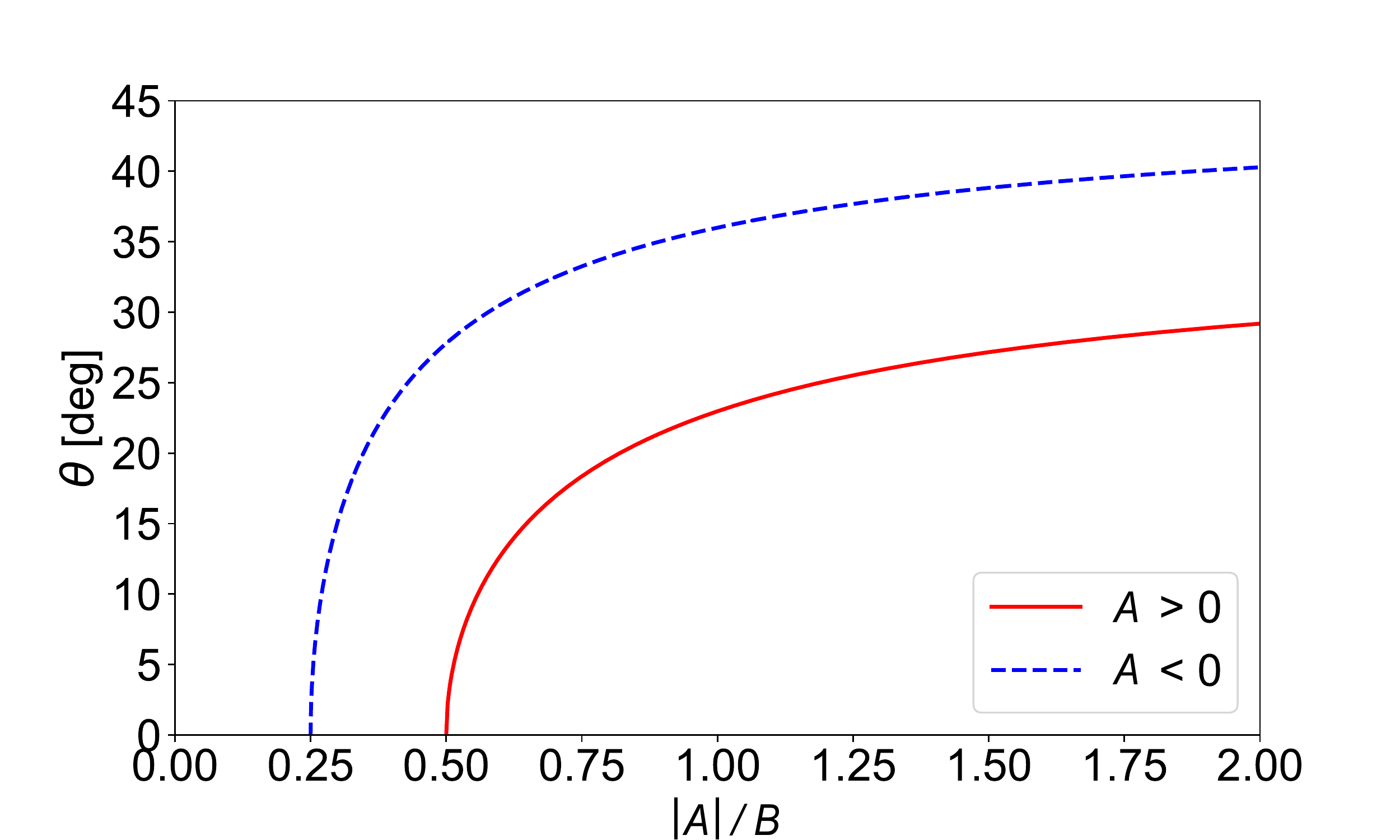}
\end{center}
\caption{
Preferred angle $\theta$ of a uniform spin configuration as a function of $|A|/B$.
For $A>0$ ($A<0$), the spins are tilted in the $x$ ($y$) direction.
When the magneto-crystalline anisotropy dominates the Zeeman energy at $A/B>0.5$ and $-A/B>0.25$, the preferred angle is tilted from the $z$ axis.}
\label{fig:prefereed_angle}
\end{figure}

In the absence of the Zeeman term, the magneto-crystalline anisotropy $U_\textrm{ii}(\SS_\rr)$ with $A>0$ is minimized when the magnetization points to one of the eight preferred directions: $\SS_\rr=(\pm 1/\sqrt{3},0,\pm \sqrt{2/3} )$ and $(\pm 1/\sqrt{3},\pm \sqrt{2/3},0)$.
An infinitesimally small Zeeman field along the $z$ axis lifts the degeneracy of these directions, and the magnetization chooses the ones having largest $z$ component, i.e., $(\pm 1/\sqrt{3},0,\sqrt{2/3})$. As the Zeeman field increases, the magnetization direction gradually changes from $(\pm  1/\sqrt{3},0,\sqrt{2/3})$ to $\bm e_z$.
Thus, the preferred direction is obtained by
assuming a uniform spin configuration
\begin{align}
\SS_\rr =\tt= (\mathrm{sin}\theta, 0, \mathrm{cos}\theta),
\label{eq:t_under_mca}
\end{align}
and minimizing the energy per spin
\begin{align}
    E(\theta)=A\left(\sin^4\theta+\frac{1}{2}\cos^4\theta\right)-B\cos\theta
\end{align}
with respect to $\theta$.
We note that since $\left.d^2E/d\theta^2\right|_{\theta=0}=-2A+B$, the energy minimum at $\theta=0$ (i.e., $\tt=\bm e_z$) for $A/B\le 0.5$ changes to a local maximum for $A/B>0.5$, and $\tt$ deviates from $\bm e_z$ at $A/B=0.5$.
We plot the numerically obtained preferred angle $\theta$ in Fig.~\ref{fig:prefereed_angle} as a function of $A/B$.
Note that $E(\theta)$ is an even function of $\theta$, which means that there are two preferred directions $\tt_\pm=(\pm \sin\theta,0,\cos\theta)$.
As we will see in Sec.~\ref{sec:numerical_phase}, the magnetic domains of $\SS_\rr=\tt_\pm$ appear in the strong anisotropy regime.

In the case of $A<0$, the magneto-crystalline anisotropy favours the magnetization direction $\SS_\rr =(\pm1,0,0)$ and $(0,\pm1,\pm1)/\sqrt{2}$, and the combination with the Zeeman term results in $\tt$ lying in the $y$-$z$ plane. The tilting angle is calculated in a similar manner as in the case of $A>0$, and the result is shown in Fig.~\ref{fig:prefereed_angle} with the dashed curve.
For $A<0$, $\theta$ becomes nonzero for $|A|/B>0.25$.

\begin{figure}
\begin{center}
\includegraphics[width=1.0\linewidth]{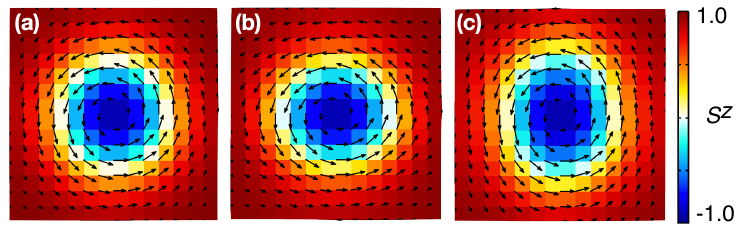}
\end{center}
\caption{Stable single-skyrmion configuration at $BJ/D^2=0.70$ and (a) $A=0$, (b) $A=0.5B$, and $A=-0.25B$, obtained as a stationary solution of the LLG equation. The size of each panel is 14 sites $\times$ 14 sites. 
The arrows indicate the vector $\SS_\rr$ projected to the $x$-$y$ plane and the color plot shows $S_\rr^z$. The circular configuration at $A=0$ (a) is elongated along the $x$ and $y$ direction for $A>0$ (b) and $A<0$ (c), respectively.}
\label{fig:Sk_shape}
\end{figure}

The anisotropy in the spin space leads to the deformation of the skyrmion configuration even when $\tt=\bm e_z$. Figure~\ref{fig:Sk_shape} shows the single skyrmion configurations for $A=0$ (a), $A=0.5B$ (b), and $A=-0.25B$ (c) at $BJ/D^2=0.70$, which clearly shows that the skyrmion for $A>0$ ($A<0$) is elongated along the $x$ ($y$) direction. Since our interest is how the inter-skyrmion interaction changes as the skyrmion deforms, it is enough to investigate only in the $A>0$ case. Although the small changes in spin configuration around the skyrmion may change the details of the interaction, the qualitative behavior is the same for both $A>0$ and $A<0$. We, therefore, discuss below only the case of $A>0$ in detail.

\subsection{Anisotropic interaction in single domain}
\label{sec:Uii_interaction}
\begin{figure}
\begin{center}
\includegraphics[width=\linewidth]{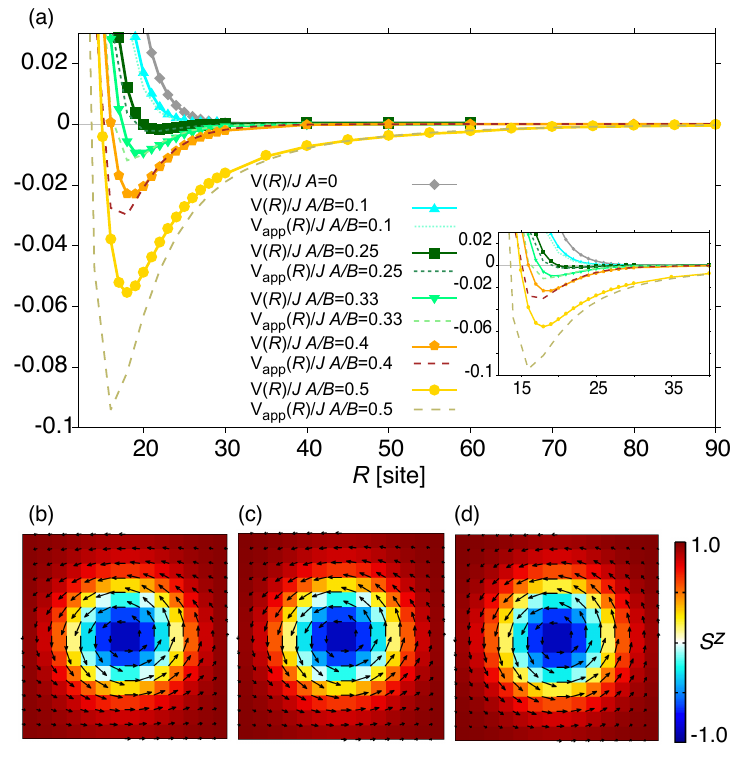}
\end{center}
\caption{
(a) 
Interaction potential $V(R)$ between two skyrmions aligned along the $x$ direction at $BJ/D^2= 0.75$ under the magneto-crystalline anisotropy $0\le A/B\le 0.5$, for which the background magnetization is $\tt=\bm e_z$. Shown are the numerically calculated interaction $V(R)$ and the approximate one $V_\textrm{app}(R)$ for each value of $A/B$.
(inset) Magnified view of $V(R)$ up to $R=40$.
(b)-(d) Stable single-skyrmion configurations at $BJ/D^2=0.75$ and (b) $A/B=0.5$, (c) $0.4$, and (d) $0.0$, obtained as a stationary solution of the LLG equation. The details of the plots are the same as those in Fig.~\ref{fig:Sk_shape}. 
Although the deformation of the skyrmion configuration in (b)-(d) is less clear than that in Figs.~\ref{fig:Sk_shape}(a) and (b), it certainly has a significant effect on the interaction potential, as shown in (a).
}
\label{fig:Vr_A1-2}
\end{figure}

Now we consider the inter-skyrmion interaction. We start from the case of $\tt=\bm e_z$.
When $A/B\lesssim 0.5$ and $B$ is moderately large, 
the background spins are not tilted but the skyrmions are well distorted due to the magneto-crystalline anisotropy.
Figure~\ref{fig:Vr_A1-2}(a) shows the interaction potential $V(R)$ of skyrmions alinged along the $x$ axis at $A/B=0.0, 0.1, 0.25, 0.33, 0.4$, and $0.5$ with $BJ/D^2=0.75$.
One can clearly see that the interaction potential becomes negative at $R\gtrsim 2R_\textrm{sk}$, and the potential becomes deeper
for larger crystalline anisotropy $A/B$.
However, the potential depth is as shallow as a few percent of $J$, which is the same order as that under a tilted magnetic field.
We also plot the approximate interaction $V_\textrm{app}(R)$ calculated from the single skyrmion solution, which agrees well with $V(R)$ up to a relatively small $R$ close to the potential minimum. For example, for the case of $A/B=0.4$, for which the interaction potential has a minimum at $R=18$ site, the two curves almost coincide with each other at $R\ge 20$.

Because $\tt=\bm e_z$ for $A/B\le 0.5$, there is only the contribution from the $J$ term to $A_{+-}$ [see Eq.~\eqref{eq:A_CM}].
Therefore, the origin of the attractive interaction is purely due to the deformation as discussed in Sec.~\ref{sec:deformation_effect}.
As shown in Figs. \ref{fig:Vr_Sk_shape}(c) and \ref{fig:Sk_shape}(b), the skyrmion  deforms such that the profile of the $x$ component, $S_\rr^x$, extends in the both $x$ and $y$ directions, which enhance the negative contribution from the $x$ component to Eq.~\eqref{eq:Vint_J_2}, resulting in the attractive interaction along the $x$ axis. 

\begin{figure}
\begin{center}
\includegraphics[width=0.65\linewidth]{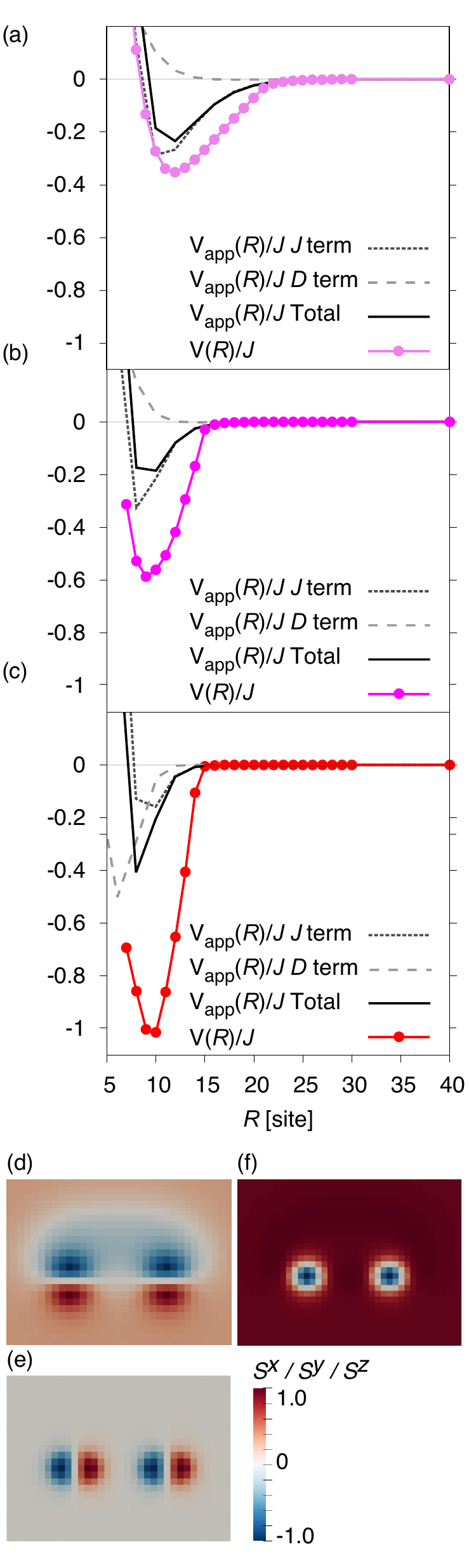}
\end{center}
\caption{
(a)-(c)
Interaction potential $V(R)$ between two skyrmions aligned along the $x$ direction under a magnetic field of strength $BJ/D^2=1.0$ and magneto-crystalline potential (a) $A/B=0.67$, (b) $1.0$, and (c) $2.0$, for which the background magnetization $\tt$ is tilted from $\bm e_z$ in the $x$ direction. Shown are the numerically calculated interaction $V(R)$, the approximate one $V_\textrm{app}(R)$, and the contributions from the $J$ and $D$ terms in Eq.~\eqref{eq:A_CM}. 
(d)-(f) Magnetization profile of a double-skyrmion configuration at $BJ/D^2=1.0$ and $A/B=1.0$ with a fixed relative position $\RR=R{\bm e_x}$ with $R=14$, where the color plots show (d) $S^x_\rr$, (e) $S^y_\rr$, and (f) $S^z_\rr$.
A small magnetic domain with $\SS_\rr^x<0$ arises between two skyrmions surrounded by the domain wall with $\SS_\rr^z=1$, which induces the strong attractive interaction shown in (a)-(c).
}
\label{fig:Vr_A1}
\end{figure}

The situation drastically changes for $A/B>0.5$.
We plot the interaction potential $V(R)$ of skyrmions aligned along the $x$ axis at $A/B=0.67, 1.0$, and $2.0$ with $BJ/D^2=1.0$ in Fig.~\ref{fig:Vr_A1}(a), (b), and (c), respectively. 
In these cases, the background spins are tilted from the $z$ axis. The approximate interaction $V_\textrm{app}(R)$ and the contributions from the $J$ term and $D$ term to $V_\textrm{app}(R)$ are also plotted in the same figure.
Differently from Fig.~\ref{fig:Vr_A1-2}, $V(R)$ in Fig.~\ref{fig:Vr_A1} becomes much stronger than $V_\textrm{app}(R)$ around the potential minimum.
The interaction energy becomes in the order of $0.1J$ to $J$ for $A/B\gtrsim 0.5$.
Although the skyrmion distance which minimizes $V(R)$ becomes smaller for Fig.~\ref{fig:Vr_A1} than that of Fig.~\ref{fig:Vr_A1-2}, we have confirmed that this is due to the difference in the value of $B$: Stronger $B$ makes the stable skyrmion distance shorter, but the minimum energy is almost insensitive to $B$.

The origin of the strong attraction along the $x$ axis is due to the formation of a magnetic domain.
Differently from the case in Sec.~\ref{sec:numerical_int_tiltB},
where $\tt$ is uniquely determined along the external magnetic field,
there are two stable uniform configurations $\tt_\pm$ in the present case.
Thus, when two skyrmions are embedded in a uniform configuration $\SS_\rr=\tt_+$, a small magnetic domain of $\SS_\rr=\tt_-$ arises between two skyrmions. We show in Figs.~\ref{fig:Vr_A1}(d)--(f) the magnetization configuration of two skyrmions located at distance $R=14$ site for $BJ/D^2=1.0$ and $A/B=1.0$. One can see that a region of $S_\rr^x<0$ is surrounded by the domain wall with $S_\rr^z=1$, and the two skyrmions are located on the domain wall. This magnetic domain strongly bounds the skyrmions.
The strong attractive interaction suggests that once the skyrmion bound-state is created, 
it will be robust against external disturbance such as thermal fluctuations.

\begin{figure}[htbp]
\begin{center}
\includegraphics[width=\linewidth]{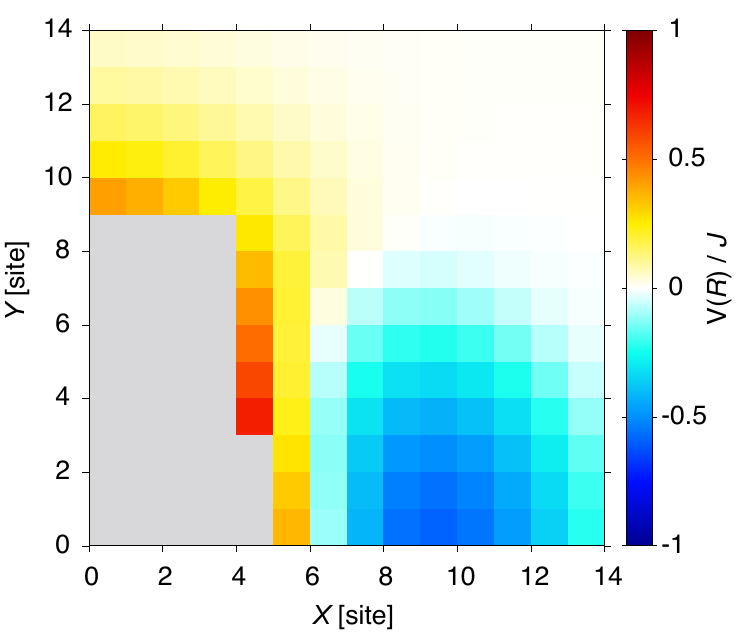}
\end{center}
\caption{
Interaction potential $V(\bm R)$ between two deformed skyrmions at relative position $\bm R=(X,Y)$ under the magneto-crystalline anisotropy $A/B=1.0$ at $BJ/D^2=1.0$.
The angular dependence of $V(\RR)$ is almost independent of the value of $A/B$,  no matter it is larger or smaller than $0.5$.
}
\label{fig:aniso_Vr}
\end{figure}

The direction dependence of the interaction potential is shown in Fig.~\ref{fig:aniso_Vr}, where $V(\bm R)$ calculated for $A/B=1.0$ and $BJ/D^2=1.0$
is plotted as a function of $\bm R=(X,Y)$.
The interaction is attractive when the angle of
the relative direction of the two skyrmions to the $x$ axis is less than $45^\circ$ and repulsive otherwise. 
We note that regardless of the exact value of $A/B$, the skyrmion's relative angle dependence of $V(\RR)$, Fig. \ref{fig:aniso_Vr}, is qualitatively the same.

\section{Novel skyrmion lattice structures due to inter-skyrmion attractions}
\label{sec:numerical_phase}

So far, we have investigated the inter-skyrmion interaction under anisotropic geometries. In this section, we discuss the ground-state structures affected by the anisotropic interaction,
focusing on the magneto-crystalline anisotropy which stabilizes magnetic domains for $A/B > 0.5$.

\begin{figure*}
\begin{center}
\includegraphics[width=\linewidth]{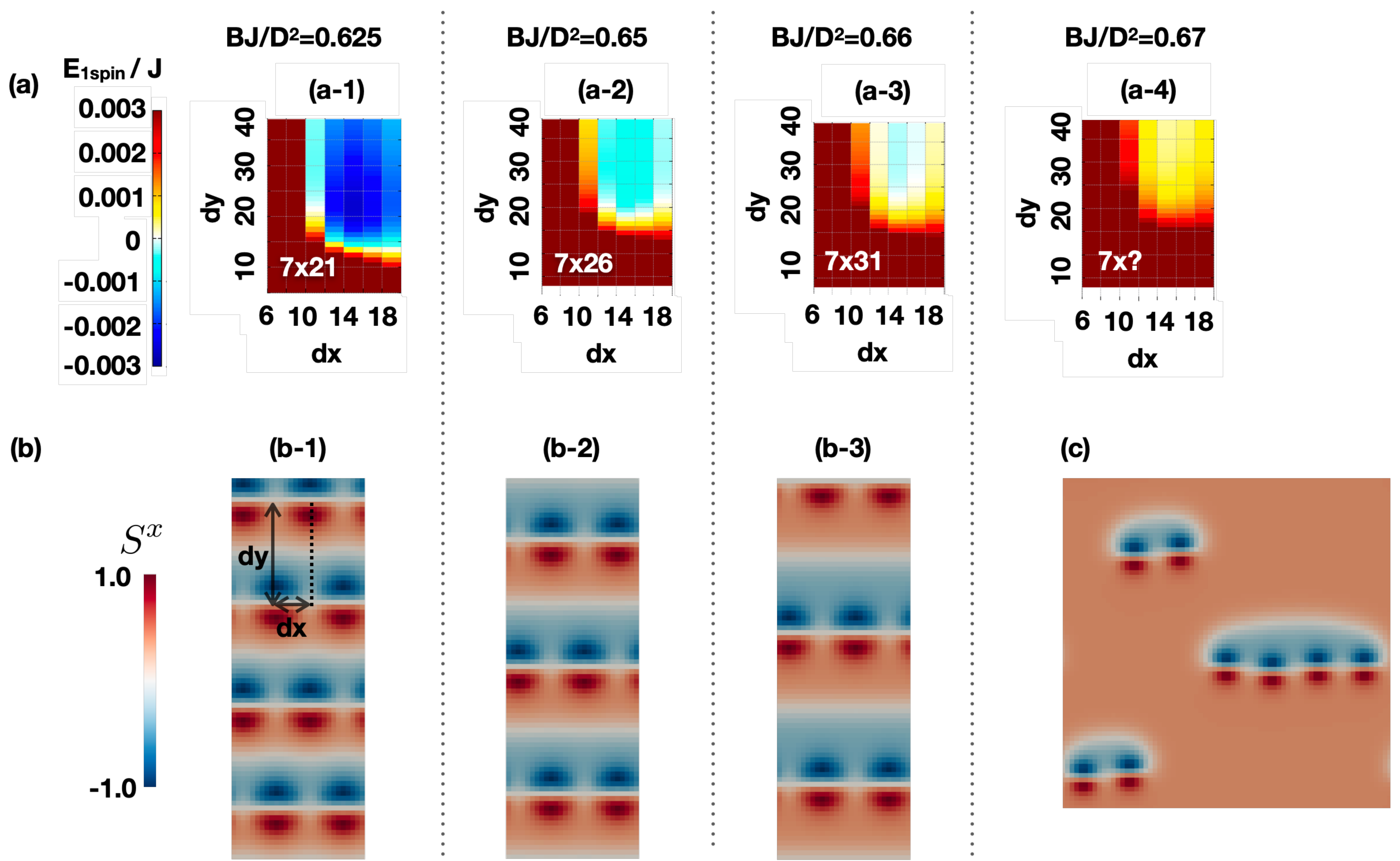}
\end{center}
\caption{
(a) Energy per spin $E_\textrm{1spin}$ for a SkX state of a periodic alignment of a unit cell of size $2dx \times 2dy$ with 2 skyrmions in it for
 (a-1) $BJ/D^2=0.625$, (a-2) $0.65$, (a-3) $0.66$, and (a-4) $0.67$. 
The white letters indicate the coordinates of the energy minimum, which are the optimal lattice spacing.
There is no minimum in (a-4).
(b) SkX structure for the optimal lattice spacing obtained in (a). Shown are the distribution of $S^x$. The size of each panel is $28$ sites $\times 78$ sites.
(c) Plot of $S^x$ at $BJ/D^2=1.0$ obtained from the MC simulations as an excited state. 
}
\label{fig:Sk_lattice_param2}
\end{figure*}

We summarize the ground-state phase diagram in the parameter space of $B$ and $A/B$ in Table~\ref{tb:phases}, 
which is obtained by the MC simulations (see Sec.~\ref{sec:method_lattice}).
In both cases of $A/B\le 0.5$ and $A/B>0.5$, there are two critical fields $B_\textrm{cr1}$ and $B_\textrm{cr2}$: We obtain a uniform spin configuration or the FM phase at $B\ge B_\textrm{cr2}$, a SkX at $B_\textrm{cr1}\le B < B_\textrm{cr2}$, and a spin helix $B<B_\textrm{cr1}$.
In the case of $A/B \le 0.5$, the magnetization in the FM phase is along the $z$ direction [Region (i) in Table~\ref{tb:phases}].
By lowering $B$ below $B_\textrm{cr2}$ with keeping $A/B \le 0.5$, a triangular SkX elongated along the $y$ axis arises [Region (ii)]. 
Here, the distortion of the triangular lattice is due to the anisotropic nature of the inter-skyrmion interaction as shown in Fig.~\ref{fig:aniso_Vr}:
The inter-skyrmion distance along the $x$ axis becomes smaller than that along the $y$ axis because of the attractive interaction along the $x$ axis.
By further lowering $B$ below $B_\textrm{cr1}$, a helical spin structure is stabilized [Region (iii)].
In the case of $A/B>0.5$, on the other hand, the FM phase has tilted magnetization $S^z\neq 1$ due to the interplay between the anisotropy potential and the Zeeman energy.
Since domain walls cost extra energy, the system favors the single domain configuration of one of the preferred directions $\tt_\pm$ [Region (iv)].
When the magnetic field becomes lower than $B_\textrm{cr2}$ [Region (v)], a triangular lattice structure elongated along the $y$ axis appears as in the case of Region (ii).
Note, however, that different from Region (ii), magnetic domains of $\SS=\tt_\pm$ appear in the background of the lattice, and skyrmions align along the domain walls [see also Fig.~\ref{fig:Sk_lattice_param2}(c)]. In Region (v), the domain walls are stabilized by accompanying skyrmions on them. 

Strictly speaking, the topological object that appears on a domain wall in Region (v) is not a skyrmion but a bimeron~\cite{Nagase-2020}. 
Here, a bimeron is a pair of merons and has the same topological charge as a skyrmion.
The difference between a skyrmion and a meron is the boundary condition on a circle surrounding the object: The magnetization direction around a skyrmion is fixed, while that around a meron winds with nonzero winding number $\pi_1(S^1)$~\cite{Gao-2019}.
In the present system, as $A/B$ increases, a skyrmion lattice changes continuously to a bimeron lattice when $\tt$ tilts from the $z$ axis. However, we here call both of them skyrmion for convenience of explanation. 

\begin{table*}[htbp]
\centering
\begingroup
\renewcommand{\arraystretch}{1.5}
\begin{tabular}{|c|c|c|} \hline
\backslashbox{$\ \ B\ \ $}{$\ \ A/B\ \ $} & $0<A/B \le 0.5$ & {$0.5 < A/B$}\\ \hline
$B_{\mathrm{cr2}} \le B$ & (i) Single Domain ($S_z = 1$) & \begin{tabular}{c}{(iv) Tilted Single Domain }{($S_z \neq 1$ and $S_x \neq 0$)}\end{tabular}\\ \hline
$B_{\mathrm{cr1}} \le B < B_{\mathrm{cr2}}$ & (ii) Elongated Triangular SkX & \begin{tabular}{c}{(v) Elongated Triangular SkX with Magnetic Domains}\end{tabular}\\ \hline
$B < B_{\mathrm{cr1}}$ & \multicolumn{2}{c|}{(iii) Helix} \\ \hline
\end{tabular}
\endgroup
\caption{Ground-state phase diagram obtained by the MC simulation under a magneto-crystalline anisotropy $A$ and an external magnetic field $B$ along the $z$ axis. 
The critical magnetic fields $B_\textrm{cr1}$ and $B_\textrm{cr2}$ are dependent on $A/B$. The single domain phases at high-field region, (i) and (iv), are the FM phase.
See text for the detailed description of the each phase.
}
\label{tb:phases}
\end{table*}

\begin{figure}
\begin{center}
\includegraphics[width=\linewidth]{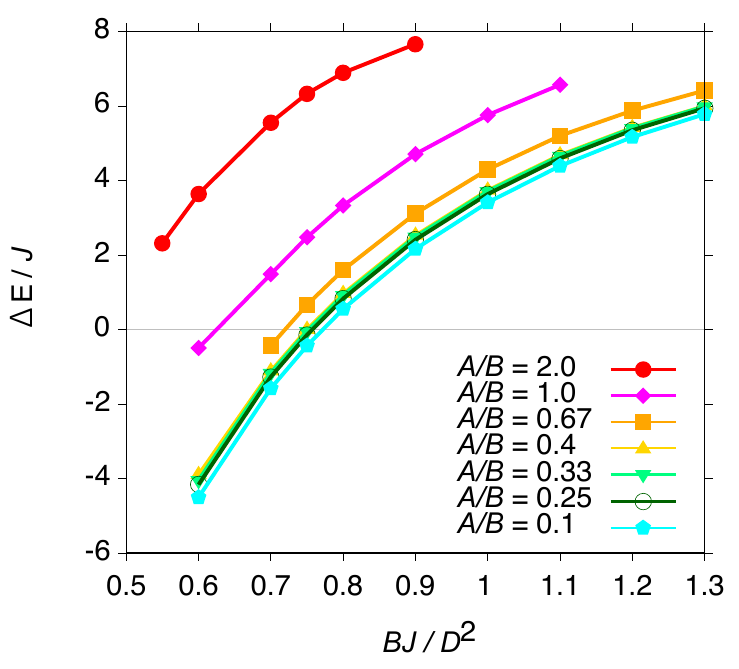}
\end{center}
\caption{
Single skyrmion energy ${\Delta E}\equiv E_\textrm{1sk}-E_\textrm{ferro}$ as a function of $BJ/D^2$ for various values of $A/B$.
}
\label{fig:ESk1}
\end{figure}

In Table~\ref{tb:phases}, the critical fields $B_\textrm{cr1}$ and $B_\textrm{cr2}$ are dependent on $A/B$.
We numerically find that $B_\textrm{cr1}$ is insensitive to the value of $A/B$ and given by $\sim 0.3D^2/J$, whereas $B_\textrm{cr2}$ is strongly dependent on $A/B$.
The latter can be explained from the $A/B$ dependence of the energy of a single skyrmion.
In the case when the inter-skyrmion interaction is always repulsive, $B_\textrm{cr2}$ is determined as the magnetic field at which the energy to create a single skyrmion in the FM state crosses zero:
Since the interaction energy between well-separated skyrmions is negligible, the skyrmion lattice becomes the ground state exactly when the single-skyrmion energy becomes negative.
In Fig.~\ref{fig:ESk1}, we show the single-skyrmion energy $\Delta E\equiv E_\mathrm{1sk}- E_\textrm{ferro}$
as a function of the strength of the external field $BJ/D^2$ for various values of $A/B$.
The strong $A/B$ dependence of the horizontal-intercept of $\Delta E$ is consistent with the $A/B$ dependence of $B_\textrm{cr2}$.
Note, however, that the inter-skyrmion interaction in the present system is attractive along the $x$ direction and hence shifts the phase boundary.

In order to investigate the phase boundary between the FM and SkX phases, we employ the LLG equation and calculate the energy of the SkX in the following manner.
We prepare a system of size $2dx \times 2dy$ as a unit cell and place skyrmions at $(dx/2,dy/2)$ and $(3dx/2,3dy/2)$ so that a periodic arrangement of this unit cell reproduces the elongated triangular lattice obtained by the MC simulation. We then calculate the energy for the stationary state in the unit cell under periodic boundary conditions.
The optimal lattice spacing $dx$ and $dy$ are determined as those minimize the energy per spin $E_\textrm{1spin}$.
Figure~\ref{fig:Sk_lattice_param2}(a) shows the $(dx,dy)$ dependence of $E_\textrm{1spin}$
for $A/B=1.0$ and $BJ/D^2=0.625$ (a-1), $0.65$ (a-2), $0.66$ (a-3), and $0.67$ (a-4).
Here, the energy is measured from that of the FM state of $\SS=\tt_+$.
The fact that the energy minimum exists and is negative in Figs.~\ref{fig:Sk_lattice_param2}(a-1)-(a-3) indicates that the SkX is the ground state at these magnetic fields.
Note that the single-skyrmion energy $\Delta E$ at $A/B=1.0$ crosses zero at $BJ/D^2=0.625$ (Fig.~\ref{fig:ESk1}),
which means that the SkX phase appearing at $BJ/D^2>0.625$ is the skyrmion condensation due to the attractive interaction.
As $B$ increases, the domain width along the $y$ direction becomes larger and larger, and eventually the domain size becomes comparable to the system size, i.e., the transition to the FM phase occurs.
In Fig.~\ref{fig:Sk_lattice_param2}(a-4), there is no energy minimum in the region of $10\le dy\le 40$, 
suggesting the phase boundary at $BJ/D^2\sim 0.66$.

Figures~\ref{fig:Sk_lattice_param2}(b-1)-(b-3) show the distribution of $S^x$ for optimal lattice spacing obtained in Figs.~\ref{fig:Sk_lattice_param2}(a-1)-(a-3), respectively.
One can clearly see that the domains of $S^x>0$ and $S^x<0$ (which corresponds to the domains of $\SS=\tt_+$ and $\tt_-$, respectively) alternately align along the $y$ direction.
Note that due to the sign of the DM interaction in our setup, the magnetization in the upper (lower) side of the skyrmion center has $S^x>0$ ($S^x<0$).
Thus, skyrmions can appear only on the domain walls where the sign of $S^x$ coincides with the skyrmion structure, and cannot exist on the other domain walls.
We also note that the optimized energies for the skyrmions located at $(dx/2,dy/2)$ and $(dx/2,3dy/2)$ in a unit cell are almost the same as those shown in Fig.~\ref{fig:Sk_lattice_param2}(a),
indicating that the inter-skyrmion interaction along the $y$ direction over the domain wall is almost negligible. 

\begin{figure}
\begin{center}
\includegraphics[width=\linewidth]{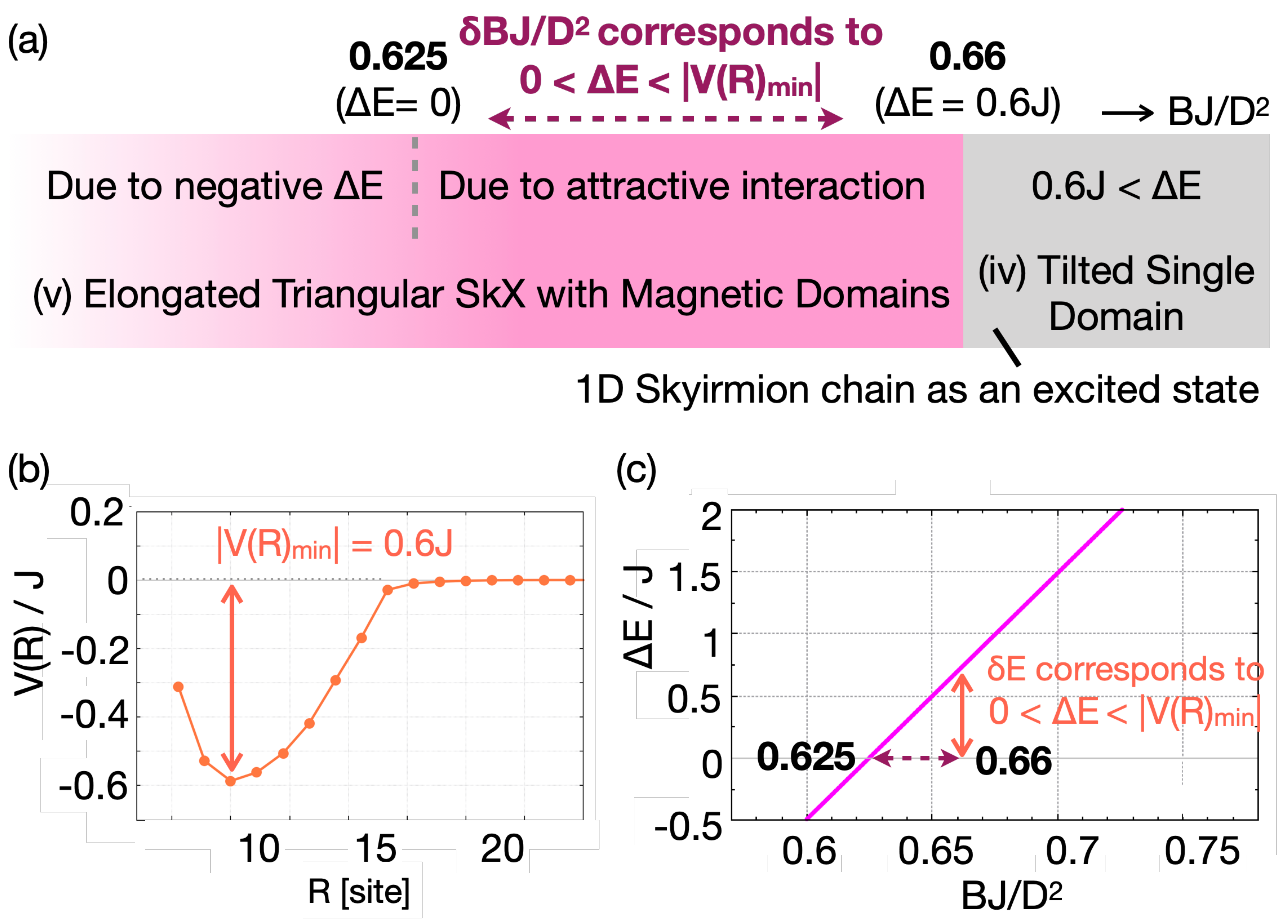}
\end{center}
\caption{
(a) Detailed ground-state phase diagram in the vicinity of the critical field $B_\textrm{cr2}$ at $A/B=1.0$. The SkX phase is divided into two regions: The one is due to the negative skyrmion energy ($BJ/D^2<0.625$), and the other is due to the attractive interaction ($0.625\le BJ/D^2< 0.66$). Above the critical field $B_\textrm{cr2}J/D^2= 0.66$, skyrmions exist as excitations and align in a 1D chain structure due to the attractive interaction along the $x$ axis.
(b) Magnified view of the inter-skyrmion interaction potential $V(R)$ in Fig.~\ref{fig:Vr_A1}(b), where the potential depth is $|V(R)_\textrm{min}|\sim 0.6J$. 
(c) Single-skyrmion energy ${\Delta E}$ as a function of $BJ/D^2$ at $A/B=1.0$, the magnified view of Fig.~\ref{fig:ESk1}. 
Solid arrow indicates $\delta E=|V(R)_\textrm{min}|$ and dotted arrow shows the corresponding $\delta B$. $\delta B$ determines the width of the SkX phase due to the attractive interaction. 
}
\label{fig:Sk_lattice_param2_2}
\end{figure}

We also note that there is an optimal $dx$ for a fixed $dy$ in Fig.~\ref{fig:Sk_lattice_param2}(a-4).
It follows that when several skyrmions are excited, they align along the $x$ axis.
Indeed, 1D skyrmion chains are obtained as a metastable state in the MC simulation as shown in Fig.~\ref{fig:Sk_lattice_param2}(c).

Finally, we summarize the detailed phase diagram for $A/B=1.0$ around $B=B_\textrm{cr2}$ in Fig.~\ref{fig:Sk_lattice_param2_2}(a).
Figures~\ref{fig:Sk_lattice_param2_2} (b) and (c) are the inter-skyrmion potential $V(R)$ at $BJ/D^2=1.0$ shown in Fig.~\ref{fig:Vr_A1}(b) and a magnified view of the single-skyrmion energy $\Delta E$ shown in Fig.~\ref{fig:ESk1},  respectively. Both are the results for $A/B=1.0$.
From Fig.~\ref{fig:Sk_lattice_param2_2}(b), one can see that the inter-skyrmion interaction energy is negative and as large as $-0.6J$ at the optimal distance.
It follows that the SkX phase is stabilized for $\Delta E\sim 0.6J$,
which corresponds to $BJ/D^2\lesssim 0.66$ as seen in Fig.~\ref{fig:Sk_lattice_param2_2}(c).
This estimation agrees well with the numerical result in Fig.~\ref{fig:Sk_lattice_param2}.
Since $\Delta E$ becomes negative for $BJ/D^2 < 0.625$ [Fig.~\ref{fig:Sk_lattice_param2_2}(a)], the SkX phase in this region is due to the negative skyrmion energy, whereas the SkX phase at $0.625\le BJ/D^2 \le 0.66$ is due to the attractive inter-skyrmion interaction.
The width $\delta B$ of the latter region is determined by the magnitude of the potential depth.
At $BJ/D^2>0.66$, the ground state is the FM phase,
which accommodates a 1D chain of skyrmions as an excitation [Fig.~\ref{fig:Sk_lattice_param2}(c)].

In the case of $A/B\le 0.5$, we obtain qualitatively the same phase diagram.
However, because the inter-skyrmion interaction is small for $A/B<0.5$,
$\delta B$ becomes much narrower than that for $A/B>0.5$.
We have also confirmed that the tilted magnetic field also gives the similar phase diagram as that of $A/B\le 0.5$, 
including the SkX phase due to the attractive interaction, and 1D skyrmion chain in an excited state.

\section{Discussion}
\label{sec:discussion}
\subsection{Dependence on crystal plane orientation}
\begin{figure}
\begin{center}
\includegraphics[width=\linewidth]{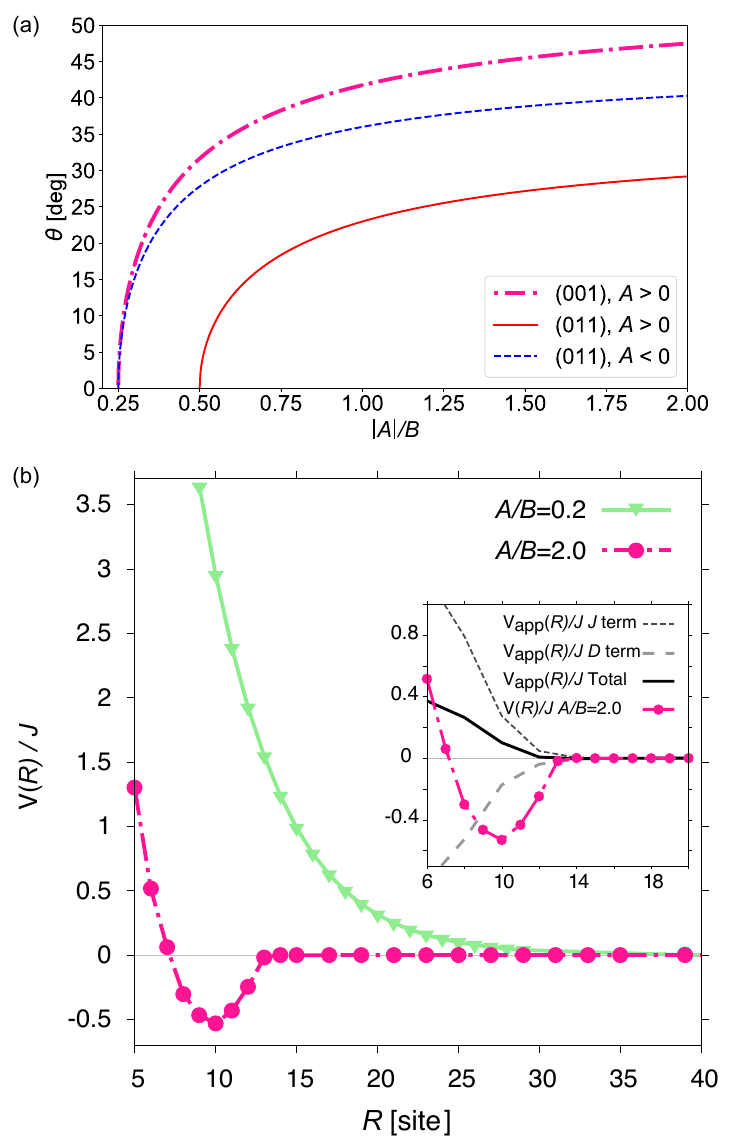}
\end{center}
\caption{
(a) Preferred angle $\theta$ of a uniform spin configuration on a (001) film as a function of $|A|/B$. 
Those for (011) film (Fig.~\ref{fig:prefereed_angle}) are shown as guides for the eye.
For $A<0$, the spins on a (001) film are parallel to $\bm e_z$ independent of $|A|/B$,
whereas they are tilted from $\bm e_z$ to the [110] direction
when the magneto-crystalline anisotropy dominates the Zeeman energy at $A/B>0.25$.
(b) Interaction potential $V(R)$ between two skyrmions aligned along the $x$ direction under a magnetic field of strength $BJ/D^2=0.75$ and magneto-crystalline potential $A/B=2.0$ and $0.2$. The background magnetization $\tt$ is tilted from $\bm e_z$ in the [110] direction for $A/B=2.0$ and parallel to $\bm e_z$ for $A/B=0.2$. 
Inset shows $V(R)$ at $A/B=2.0$ in the main panel, the approximate one $V_\textrm{app}(R)$, 
and the contributions from the $J$ and $D$ terms in Eq.~\eqref{eq:A_CM}. 
}
\label{fig:Vr_not_rotated_A}
\end{figure}

We have discussed the magneto-crystalline anisotropy in a (011) thin film in which the $C_4$ symmetry breaks.
Here, we discuss how the above results change in a (001) thin film which preserves the $C_4$ symmetry.
The magneto-crystalline anisotropy potential in a (001) thin film, which is given by $A\left[(S^x_\rr)^4+{(S^y_\rr)^4}+{(S^z_\rr)^4}\right]$,
has eight easy axes along $\langle 111\rangle$ directions for $A>0$
and six easy axes along $\langle 100\rangle$ directions for $A<0$.
In the case of $A<0$, the Zeeman field along the $z$ direction lifts the degeneracy of the easy axes, and the magnetization direction in a uniform solution is uniquely determined to be [001].
Hence, no domain structure appears for $A<0$.
On the other hand, in the case of $A>0$, the system under a Zeeman field along the $z$ axis favors a magnetization direction between the $z$ axis and [111] direction, or the other three equivalent directions $[\bar{1}11], [1\bar{1}1]$, and $[\bar{1}\bar{1}1]$.
Taking the angle from the $z$ axis as $\theta$, we minimize the anisotropy potential and obtain the optimal $\theta$, as in the case of the (011) film.
As shown in Fig.~\ref{fig:Vr_not_rotated_A}(a), $\theta$ becomes nonzero for $A/B>0.25$, suggesting the strong attractive interaction in this region.

Figure~\ref{fig:Vr_not_rotated_A}(b) shows the inter-skyrmion potential $V(\RR=R\bm e_x)$ under a perpendicular ($A/B=0.2$) and tilted ($A/B=2.0$) background uniform magnetization.
One can see that the interaction energy for $A/B=0.2$ is positive for all $R$.
This is because the shape of a skyrmion is almost undistorted owing to the $C_4$ symmetry.
For the case of $A/B=2.0$, on the other hand, the interaction potential has large negative minimum, which originates from the domain formation between the skyrmions, as in the case of $A/B>1.0$ in a (011) film.
However,  $|V(r)_\textrm{min}|$ is small compared with Fig.~\ref{fig:Vr_A1}(c).
This is again due to the $C_4$ symmetry of the system.
In the inset of Fig.~\ref{fig:Vr_not_rotated_A}(b), we show the approximate interaction $V_\textrm{app}(R\bm e_x)$ and the contributions from the $J$ and $D$ terms to it.
One can see that the contribution from the $J$ term is positive, and 
as a whole the skyrmion deformation does not enhance the attractive interaction.

\begin{figure}
\begin{center}
\includegraphics[width=\linewidth]{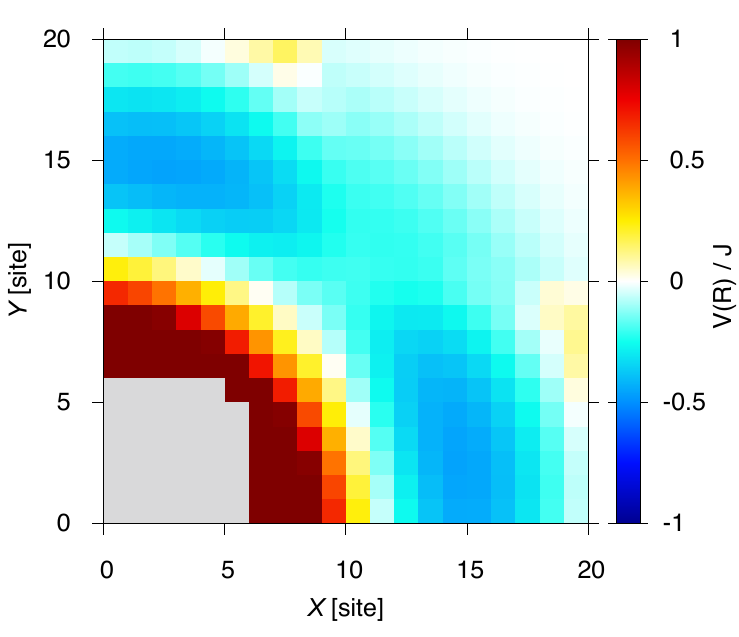}
\end{center}
\caption{
Interaction potential $V(\bm R)$ between two skyrmions at relative position $\bm R=(X,Y)$ on a (001) thin film
with the magneto-crystalline anisotropy $A/B=1.0$ at $BJ/D^2=0.75$. The potential $V(\bm R)$ preserves the $C_4$ symmetry of the crystal.
}
\label{fig:Vr_aniso_not_rotated_A}
\end{figure}

The direction dependence of the interaction potential is shown in 
Fig.~\ref{fig:Vr_aniso_not_rotated_A}, which clearly reflects the $C_4$ symmetry of the system.
Though the maximum strength of the attractive interaction is slightly smaller than the (011) film,
the attractive coupling can be found for all the relative angle direction.
This is caused by the fact that there are 4 types of magnetic domains (i.e., 4 magnetization directions preferred in a uniform solution), and
the magnetic domain arises between two skyrmions aligned 
either along the $x$ or $y$ directions.

As for the ground-state phase diagram, there are two differences from that of $(011)$ film.
 First, the SkX phase due to the attractive interaction arises only for nonzero $\theta$ at $A/B>0.25$. Second, because of the $C_4$ symmetry, a square lattice of skyrmions becomes stable in the intermediate magnetic field region, and the four magnetic domains alternatively align.

\subsection{Combined effect of in-plane magnetic field and magneto-crystalline anisotropy}
\begin{figure}
\begin{center}
\includegraphics[width=\linewidth]{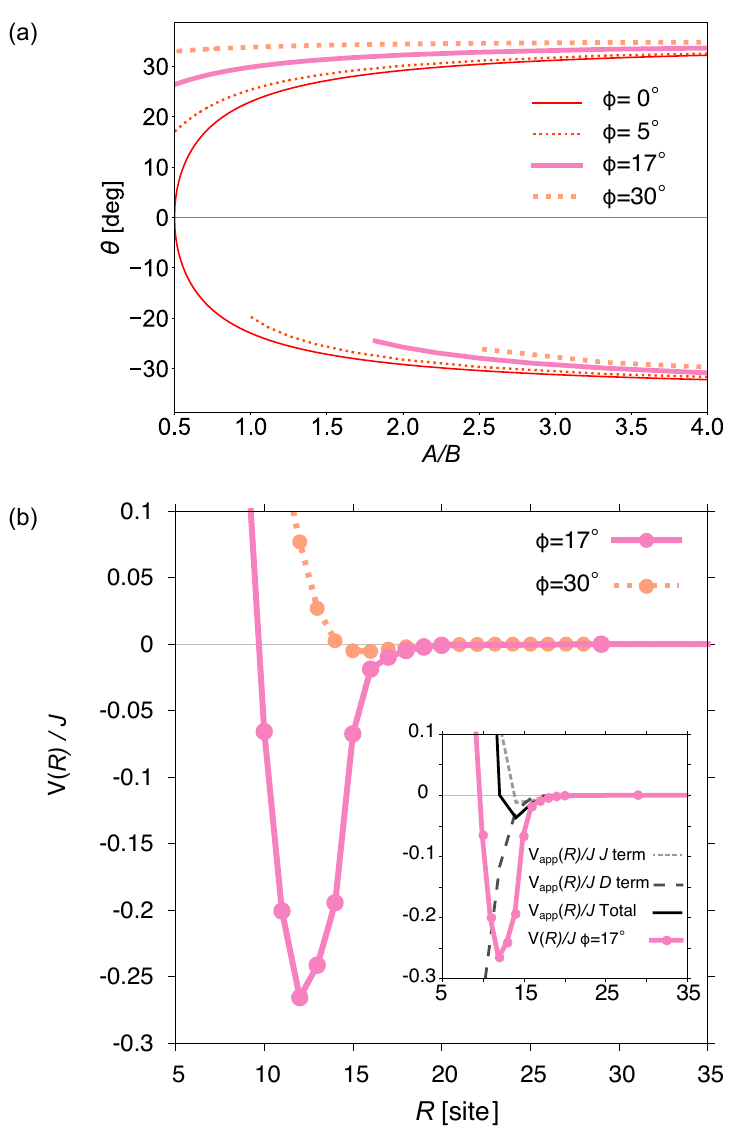}
\end{center}
\caption{
(a) Preferred angle $\theta$ of a uniform spin configuration in the presence of a in-plane magnetic field along the $x$ direction and the magneto-crystalline anisotropy $A>0$. Shown are stable ($\theta>0$) and metastable ($\theta<0$) solutions as a function of $A/B$, where $\theta$ and $\phi$ are the direction of the spins and the external magnetic fields from the $z$ axis in the $x$-$z$ plane.
The metastable solution appears above a certain threshold of $A/B$.
(b) Interaction potential $V(R)$ between two skyrmions aligned along the $x$ direction at $BJ/D^2=0.5$, $A/B=2.0$, and $\phi=17^\circ$ and $30^\circ$.
In the both cases of $\phi=17^\circ$ and $30^\circ$, the background magnetization $\tt$ is tilted from $z$ direction.
Since the metastable solution $\theta<0$ exists only for $\phi=17^\circ$,
the magnetic domain of negative $\theta$ is formed between the skyrmions, resulting in the large attractive potential.
On the other hand, for the case of $\phi=30^\circ$, the interaction potential is quite shallow due to the absence of the metastable domain.
Inset shows the one at $\phi=17^\circ$ in the main panel, together with the approximate potential $V_\textrm{app}(R)$, 
and the contributions from the $J$ and $D$ terms in Eq.~\eqref{eq:A_CM}. 
}
\label{fig:Vr_tiltB_and_A}
\end{figure}

Next, we consider the combined effect of the in-plane magnetic field and the magneto-crystalline anisotropy (the $A$ term) in a (011) thin film.
Here, we use the anisotropy potential
\begin{align}
    U_\textrm{iii}(\SS_\rr)=&-B(S^z_\rr\cos\phi+S^x_\rr\sin\phi)\nonumber\\
    &+A\left[(S^x_\rr)^4+\frac{(S^y_\rr+S^z_\rr)^4}{4}+\frac{(-S^y_\rr+S^z_\rr)^4}{4}\right],
\end{align}
for which the stable uniform configuration is written in the same form as Eq.~\eqref{eq:t_under_mca}.
However, the in-plane magnetic field resolves the degeneracy of 
the two preferred direction $\pm\theta$ in Fig.~\ref{fig:prefereed_angle}.
Figure~\ref{fig:Vr_tiltB_and_A}(a) shows the angle $\theta$ of a stable and metastable solutions for various $\phi$.
Here, the positive (negative) $\theta$ is for the stable (metastable) solution $\SS_\rr=\tt_+$ ($\SS_\rr=\tt_-$),
and the metastable solution disappears for small $A/B$.
It follows that when two skyrmions are in the background magnetization $\SS_\rr=\tt_+$, they strongly interact with each other by creating a magnetic domain of $\tt_-$ between them, if the metastable solution $\tt_-$ exists.

We can indeed see the significant change of the strength of the attractive interaction $V(\RR=R{\bm e}_x)$ depending on whether the metastable magnetic domain exists, as shown in Fig.~\ref{fig:Vr_tiltB_and_A}(b). 
In Fig.~\ref{fig:Vr_tiltB_and_A}(b), we plot $V(\RR=R\bm e_x)$ for $\phi=17^\circ$ and $30^\circ$ at $A/B=2.0$.
According to Fig.~\ref{fig:Vr_tiltB_and_A}(a), a metastable state exists (does not exist) for $\phi=17^\circ$ ($30^\circ$).
Correspondingly, the interaction potential has a deep (shallow) well for $\phi=17^\circ$ ($30^\circ$).
The inset in Fig.~\ref{fig:Vr_tiltB_and_A}(b) shows that the large attractive interaction originates from the formation of the domain, 
since the approximate potential $V_\mathrm{app}(R)$ cannot reproduce $V(R)$.

We note that the potential depth for $\phi=17^\circ$ is shallower than that of Fig.~\ref{fig:Vr_A1}(c)
and the depth for $\phi=30^\circ$ is shallower than that of Fig.~\ref{fig:tiltB_Vr}(c).
The former is because the domain of $\tt_-$, appearing between two skyrmions, has larger anisotropy potential than that of the background magnetization, i.e., $U_\textrm{iii}(\tt_-) > U_\textrm{iii}(\tt_+)$, whereas they are degenerate for $\phi=0$.
On the other hand, the latter is due to the ways of skyrmion deformation:
Under an in-plane magnetic field along the $x$ axis, the area of $S_x<0$ ($S_x>0$) becomes smaller (larger) than the case of $\phi=0$,
which gives a smaller negative contribution to Eq.~\eqref{eq:Vint_J_2}
[see Fig.~\ref{fig:Vr_Sk_shape}(b)].

\begin{figure}
\begin{center}
\includegraphics[width=\linewidth]{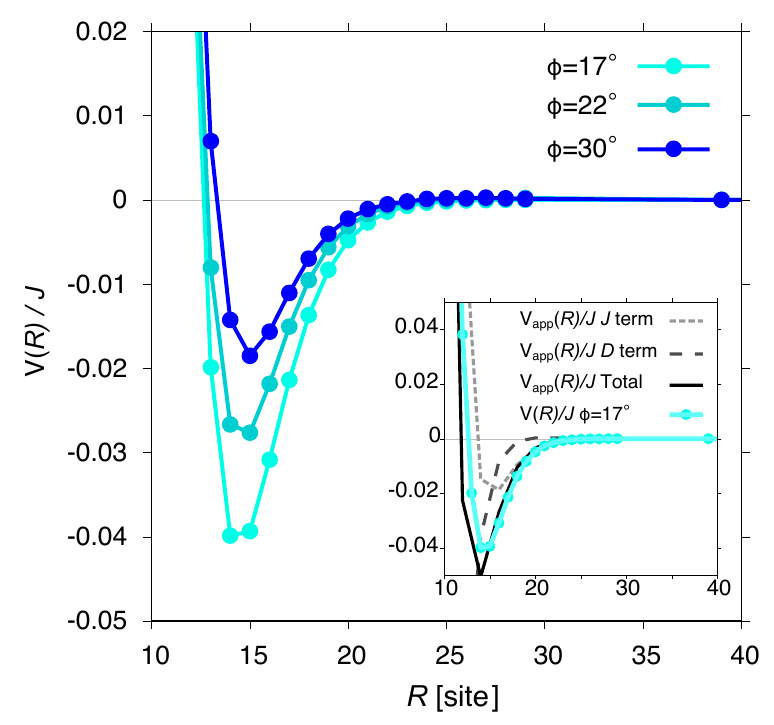}
\end{center}
\caption{
Interaction potential $V(R)$ between two skyrmions aligned along the $x$ direction at $A/B=0.5$, $BJ/D^2=0.75$ and  $\phi=17^\circ, 22^\circ$, and $30^\circ$. 
Inset compares the data at $\phi=17^\circ$ with the approximate one $V_\textrm{app}(R)$, and the contributions from the $J$ and $D$ terms in Eq.~\eqref{eq:A_CM}.
}
\label{fig:Vr_tiltB_and_A-2}
\end{figure}

This deformation also changes the $\phi$ dependence of the inter-skyrmion interaction.
Figure~\ref{fig:Vr_tiltB_and_A-2} shows the inter-skyrmion interaction potential for various $\phi$ at $A/B=0.5$, where no domain is formed between the skyrmions. 
As $\phi$ increases from zero, the potential well becomes deeper first, but it becomes shallower for larger $\phi$. 
This behavior differs from the case of $A=0$ (see Fig.~\ref{fig:tiltB_Vr}), where the potential depth monotonically increases as a function of $\phi$.
The inset confirms that the attractive interaction is mainly from the $J$ term, i.e., the distortion of the skyrmions.
We note that the minimum energy is almost insensitive to $B$.

\subsection{Realistic values of magneto-crystalline anisotropy $A$}
From our calculations, $|A|/B_{\mathrm{cr2}}\ge 0.5$ is required to observe 
the large attractive interaction and the domain wall skyrmions.
The observed values in real materials for the ratio are $|A|/B_{\mathrm{cr2}}\sim 0.00364$ in a Cu$_2$OSeO$_3$ thin film at 5~K\cite{Seki-2012, Zhang-2016, Stasinopoulos-2017}, 
$|A|/B_{\mathrm{cr2}}\sim 0.385$ in a Fe$_{0.7}$Co$_{0.3}$Si thin film at 5~K \cite{Shimizu-1990, Porter-2013}, and $|A|/B_{\mathrm{cr2}}\sim 1.59$ in a Co$_{8.5}$Zn$_{7.5}$Mn$_4$ thin film at 330~K \cite{Nagase-2019, Nagase-2020}.
In the last material, the domain wall skyrmions (or bimerons) indeed appear in a thin film with the thickness $\sim 50$~nm.

\subsection{Bound states at finite temperature}
One might wonder how relevant the inter-skyrmion interaction obtained at 0~K at finite temperature is.
We note that the binding energy, $|V(\RR)_\textrm{min}|$, becomes as large as $J$, which is the same order of the energy of a skyrmion:
In the presence of a single skyrmion in a FM state, the contributions of the $J$ and $D$ terms in the energy functional~\eqref{eq:energy_functional} is evaluated as $\pi J/2$ by approximating $|\bm\nabla \bm n|^2\sim 1/R_\textrm{sk}^2$ and $\bm \nabla\times \nn\sim -1/R_\textrm{sk}$ only inside the area $\pi R_\textrm{sk}^2$
with $R_\textrm{sk}\sim Ja/D$.
 If we evaluate the energy per spin, these energies are quite small compared with, e.g., the spin-exchange interaction, because a skyrmion involves so many spins.
However, given that skyrmions are visible under large thermal fluctuations at room temperature, skyrmion-bound states with comparable binding energies would be observable in the same temperature range.

\section{Conclusion}
\label{sec:conclusion}
In conclusion, we have shown that in-plane anisotropy in 2D chiral magnets can induce inter-skyrmion attractions via deforming a skyrmion shape or creating a magnetic domain between skyrmions.
We have investigated inter-skyrmion interactions and stable spin configurations in 2D chiral magnets under a tilted magnetic field and/or with the magneto-crystalline anisotropy on a (011) thin film.
We first describe the approximate inter-skyrmion interaction $V_\mathrm{app}(R)$ at a large distance in terms of a single skyrmion configuration, using which we qualitatively explain that the deformation of a skyrmion shape can change the sign of the interaction.
Our numerical calculations exhibit that the inter-skyrmion interaction $V(R)$ under an anisotropic geometry is indeed a weak attraction in a certain direction and agrees with $V_\mathrm{app}(R)$.
However, when the magneto-crystalline anisotropy is sufficiently large,
the inter-skyrmion attraction becomes much stronger than that expected from $V_\textrm{app}(R)$.
Such a large attractive interaction, $\sim J$, is attributed to the formation of a magnetic domain between the two skyrmions.
In the ground state, the inter-skyrmion attraction stabilizes the SkX and enhances the upper critical magnetic field of the SkX phase.
Under a strong magneto-crystalline anisotropy, 1D alignments of strongly bound skyrmions form domain walls, which in turn are aligned to form an elongated triangular lattice of bimerons with magnetic domains in its background.
A 1D chain of tightly bound skyrmions also exists in the FM phase as an excitation.

We further demonstrate that the angular dependence of the inter-skyrmion interaction strongly depends on the crystal plane orientation to the 2D film. For example, the strong magneto-crystalline anisotropy on a (001) film induces attractive interaction in every direction, whereas the interaction on a (011) film is attractive along the $x$ axis and repulsive along the $y$ asis.
We also investigate the combined effect of the in-plane magnetic field and the magneto-crystalline anisotropy on a (011) film and find that the magnitude of the attraction is tunable in a wide range via controlling the formation of magnetic domains by changing the direction of the external magnetic field.
Such high controllability of inter-skyrmion interactions will open further possibilities for utilizing skyrmions as an information medium. 

In this work, we have neglected the anisotropic exchange interactions [the $K$ term in Eq.~\eqref{eq:Uii}], the magnetic dipole-dipole interactions, and the 3D configuration in a film with finite width. Although they are crucial for a quantitative evaluation of the inter-skyrmion interaction in actual materials, we leave the detailed investigation of these effects as a future issue.
As for the effect of 3D configuration, the experimentally observed attraction between skyrmions modulated along the $z$ direction~\cite{Loudon-2018,Du-2018} can be explained by the mechanism we have found: 
The magnetization configuration in a 2D cross section of the 3D system has a tilted background magnetization, similar to the case under a tilted magnetic field, and skyrmions are no more circularly symmetric; Applying our result to the 2D cross section, the sign of the inter-skyrmion interaction depends on the direction of the background magnetization;
Stacking such 2D planes along the z direction, on average, results in an attractive interaction.
Similarly, modulation of skyrmion shape in time, due to thermal fluctuations or by an external control, is expected to modify the inter-skyrmion interaction. 
The result of this paper would give a guiding principle for designing such an effective inter-skyrmion interaction, which is our future interest.

\section{Acknowledgement}
We would like to thank M. Nagao, T. Nagase, X. Z. Yu, W. Koshibae, M. Mochizuki, and J. Barker for fruitful discussions and suggestions.
This work was supported by JST-CREST (Grant No. JPMJCR16F2), JSPS KAKENHI
(Grants No. JP18K03538 and No. JP19H01824), and Toyota Riken Scholar.
M. K. was supported by Grant-in-Aid for JSPS Fellows (JP19J20118) and GP-Spin at Tohoku University.

\appendix
\section{Composite skyrmion state}
\label{sec:app_stereographic}
We introduce a procedure to construct a two-skyrmion state from a single-skyrmion state.
First, we define the stereo-graphic projection
\begin{align}
    p:\mathbb{C}\cup \infty \mapsto S^2,
\end{align}
that maps a complex number $u=u_1+iu_2$ ($u_1,u_2\in\mathbb{R}$) to a three-dimensional unit vector:
\begin{align}
    p(u)=\left(\frac{2u_1}{1+|u|^2},\frac{2u_2}{1+|u|^2},\frac{1-|u|^2}{1+|u|^2}\right).
\label{eq:stereographic}
\end{align}
The projection $p$ maps $u=0$ to $p(0)=(0,0,1)$ and $|u|=\infty$ to $p(\infty)=(0,0,-1)$.
We also introduce an orthogonal transformation $\mathcal{R}$ as a rotation about ${\bm e}_z\times \tt$ by an angle $\cos^{-1}({\bm e}_z\cdot\tt)$, which satisfies $\mathcal{R}({\bm e}_z)=\tt$. The combined operator $p_\tt(u)\equiv \mathcal{R}[p(u)]$
maps $u=0$ to $p_\tt(0)=\tt$.
For a skymion filed $\nn(\rr)$ that satisfies $\nn(\infty)=\tt$, the boundary condition in terms of $u(\rr)=p_\tt^{-1}[\nn(\rr)]$ is given by $u(\infty)=0$.

Suppose that we have two single-skyrmion solutions $\nn_u(\rr)=p_{\tt}(u)$ and $\nn_v(\rr)=p_{\tt}(v)$, which have concentrated skyrmion charge densities $\rho_{\rm ch}(\rr)=\nn\cdot(\partial_x\nn\times \partial_y\nn)/(4\pi)$ at around $\rr=\rr_u$ and $\rr_v$, respectively.
Then, a composite skyrmion states is given by
\begin{align}
    \nn_w(\rr)=p_\tt (w), \ \ w=u+v.
    \label{eq:combined_skyrmion}
\end{align}
This procedure preserves the total skyrmion charge with keeping the boundary condition:
If $\nn_u(\infty)=\nn_v(\infty)=\tt$,  i.e., $u(\infty)=v(\infty)=0$, we obtain $\nn_w(\infty)=p_\tt(0+0)=\tt$.
Equation~\eqref{eq:combined_skyrmion} well describes the composite skyrmion state
when the distance $|\rr_u-\rr_v|$ is large enough.

We approximate $\nn_w(\rr)$ using $\nn_{u,v}(\rr)$.
In the discussion below in this section, we assume $\tt=\bm e_z$ for the sake of simplicity. The result for a general $\tt$ is given by applying $\mathcal{R}$ for all $\nn_u, \nn_v$, and $\nn_w$.
As a general property of unit vector fields, when a unit vector $\nn$ is close to $\nn_0$, we can expand $\nn$ as
\begin{align}
    \nn \simeq \nn_0 + \be\times \nn_0 + \frac{1}{2}\be \times \left(\be \times \nn_0\right),
\label{eq:expansion_unitvector}
\end{align}
where $\be$ is defined such that $\be\cdot\nn_0=0$.
In Eq.~\eqref{eq:expansion_unitvector}, the normalization condition for $\nn$ is satisfied up to the second order of $\be$:
\begin{align}
    |\nn|^2=1+o(\be^2).
\end{align}
Defining $\delta\nn=\be\times \nn_0$, we can express $\be$ as
\begin{align}
    \be=\tt\times\delta \nn.
\end{align}
For example, since $\nn_{a=u,v}(\rr)$ far from the skyrmion center $\rr=\rr_a$ is close to $\tt$, we can expand it as
\begin{align}
    \nn_a\simeq \tt+\be_a^0\times\tt + \frac{1}{2}\be_a^0\times \left(\be_a^0\times \tt\right), \ \ a=u,v
    \label{eq:approx_na}
\end{align}
When $\tt=\bm e_z$ as we assumed in the above, the expansion of Eq.~\eqref{eq:stereographic} around $u=0$ gives
\begin{align}
    \delta \nn_u&=(2u_1,2u_2,0),\\
    \be_u^0&=\tt\times \delta \nn_u=(-2u_2,2u_1,0).
\end{align}
Similarly, we obtain $\delta \nn_v=(2v_1,2v_2,0)$ and $\be_v^0=\tt\times \delta \nn_v=(-2v_2,2v_1,0)$.

Using the above equations,
the combined configuration $\nn_w$ is approximated as follows. 
When $|\rr_u-\rr_v|$ is large enough, $v=p^{-1}(\nn_v)$ satisfies $|v|\ll 1$ at around $\rr=\rr_u$.
Thus, we can expand $\nn_w$ up to the linear terms of $v$ as
\begin{align}
    \nn_w=&p(u+v)\nonumber\\
    =&\left(\frac{2(u_1+v_1)}{1+|u+v|^2},\frac{2(u_2+v_2)}{1+|u+v|^2},\frac{1-|u+v|^2}{1+|u+v|^2}\right)\nonumber\\
    =&\nn_u +\left(\frac{2v_1}{1+|u|^2},\frac{2v_2}{1+|u|^2},-\frac{2(u_1v_1+u_2v_2)}{1+|u|^2}\right)\nonumber\\
    &-\frac{2(u_1v_1+u_2v_2)}{1+|u|^2}\nn_u+O(v^2)\nonumber\\
    =&\nn_u+\frac{1}{2}\left[(1+n_{uz})\delta\nn_v - (\nn_u\cdot\delta\nn_v)(\bm e_z+\nn_u)\right]\nonumber\\
    &+O(v^2).
\end{align}
Using Eq.~\eqref{eq:expansion_unitvector}, $\nn_w$ up to the second order of $v$ is given by
\begin{align}
    \nn_w &= \nn_u + \be_v\times \nn_u + \frac{1}{2}\be_v\times(\be_v\times \nn_u)+O(v^3),\label{eq:nw_expand_around_ru}\\
    \be_v &= \frac{1}{2} \nn_u\times \left[(1+\nn_u\cdot \bm e_z)\delta\nn_v-(\nn_u\cdot\delta \nn_v)\bm e_z\right].
    \label{eq:be_v}
\end{align}
For the case of $\tt\neq \bm e_z$, we apply $\mathcal{R}$ for all vector fields, obtaining Eq.~\eqref{eq:nw_expand_around_ru} with 
\begin{align}
    \be_v &= \frac{1}{2} \nn_u\times \left[(1+\nn_u\cdot \tt)\delta\nn_v-(\nn_u\cdot\delta \nn_v)\tt\right].
\end{align}
Similarly, we can expand $\nn_w$ around $\rr=\rr_v$ as
\begin{align}
    \nn_w &= \nn_v + \be_u\times \nn_v + \frac{1}{2}\be_u\times(\be_u\times \nn_v)+O(u^3),\label{eq:nw_expand_around_rv}\\
    \be_u &= \frac{1}{2} \nn_v\times \left[(1+\nn_v\cdot \tt)\delta\nn_u-(\nn_v\cdot\delta \nn_u)\tt\right].
\end{align}

\section{Derivation of $V_\textrm{app}(\RR)$}
\label{sec:app_interaction}
We first derive the equation that a stationary solution under the energy functional~\eqref{eq:energy_functional} satisfies.
Suppose that $\nn_0(\rr)$ is a stationary solution satisfying  the boundary condition $\nn_0(\infty)=\tt$.
Using Eq.~\eqref{eq:expansion_unitvector}, 
a magnetization configuration with a small fluctuation around $\nn_0(\rr)$ can be described as
\begin{align}
\nn(\rr)\simeq \nn_0(\rr)+\be(\rr)\times\nn_0(\rr),
\end{align}
up to the first order of $\be(\rr)$.
The energy difference between the configurations of $\nn(\rr)$ and $\nn_0(\rr)$ is given by
\begin{align}
&F[\nn(\rr)]-F[\nn_0(\rr)]\nonumber\\
&\simeq \int \frac{d^2r}{a^2}\left[ \frac{\partial f(\nn_0)}{\partial \nn}\cdot(\be\times \nn_0) 
+ \frac{\partial f(\nn_0)}{\partial \partial_i\nn}\cdot \partial_i (\be\times \nn_0)\right]\nonumber\\
&=\int \frac{d^2r}{a^2}\left[ \frac{\partial f(\nn_0)}{\partial \nn} - \partial_i \frac{\partial f(\nn_0)}{\partial \partial_i\nn}\right]\cdot
(\be\times \nn_0) \nonumber\\
&=\int \frac{d^2r}{a^2}\be \cdot \left\{\nn_0\times \left[ \frac{\partial f(\nn_0)}{\partial \nn} - \partial_i \frac{\partial f(\nn_0)}{\partial \partial_i\nn}\right]\right\},
\end{align}
from which the stationary solution $\nn_0(\rr)$ should satisfies
\begin{align}
\nn_0\times \left[ \frac{\partial f(\nn_0)}{\partial \nn} - \partial_i \frac{\partial f(\nn_0)}{\partial \partial_i\nn}\right]=0.
\label{eq:stationary_condition}
\end{align}
This equation is equivalent to the condition for a stationary solution of the LLG equation, $d\nn/dt=-\nn\times \bm B_{\rm eff}+\alpha \nn\times d\nn/dt$, 
with the effective magnetic field  
\begin{align}
\bm B_{\rm eff}= -\frac{\partial f(\nn)}{\partial \nn} + \partial_i \frac{\partial f(\nn)}{\partial \partial_i\nn}.
\end{align}

Now, we consider the interaction between skyrmions located at $\rr=\pm\RR/2$.
Suppose that a single-skyrmion solution with a skyrmion at $\rr=0$ is given by $\nn_\textrm{1sk}(\rr)$, which satisfies Eq.~\eqref{eq:stationary_condition}. The single-skyrmion state with a skyrmion at $\rr_u=\RR/2$ and $\rr_v=-\RR/2$ are given by $\nn_u(\rr)=\nn_\textrm{1sk}(\rr- \RR/2)$ and $\nn_v(\rr)=\nn_\textrm{1sk}(\rr+ \RR/2)$,
whereas the double-skyrmion state is approximated by the composite skyrmion state introduced in Sec.~\ref{sec:app_stereographic}:  $\nn_\textrm{2sk}\simeq\nn_w=p_\tt(p_\tt^{-1}(\nn_u)+p_\tt^{-1}(\nn_v))$.

We derive an approximate form of Eq.~\eqref{eq:Vint} at large $|\RR|$.
We divide the region of the integral into $D_+$ and $D_-$, which are the right and left sides of $\Gamma$ in Fig.~\ref{fig:config}, respectively, and rewrite Eq.~\eqref{eq:Vint} as
\begin{align}
V_{\rm int}(\RR)&=V_+(\RR)+V_-(\RR),\label{eq:Vint=V++V-}\\
V_\pm(\RR)&\equiv\int_{D_\pm} \frac{d^2r}{a^2}[f(\nn_\textrm{2sk})-f(\nn_u)-f(\nn_v)+f(\tt)].
\end{align}
When $|\RR|$ is large enough compared with the skyrmion size, 
we can approximate $\nn_\textrm {2sk}$ with the right-hand side of Eq.~\eqref{eq:nw_expand_around_ru} and $\nn_v$ with Eq.~\eqref{eq:approx_na} in $D_+$.
We further expand the integrand up to the first order in $\be_v$ and $\be_v^0$, obtaining
\begin{widetext}
\begin{align}
V_+(\RR)&\simeq \int_{D_+}\frac{d^2 r}{a^2} \left[f\left(\nn_u + \be_v\times \nn_u+\frac{1}{2}\be_v\times(\be_v\times \nn_u)\right) - f(\nn_u) - f\left(\tt+\be_v^0\times \tt+\frac{1}{2}\be_v^0\times(\be_v^0\times \tt)\right) + f(\tt)\right]\nonumber\\
&\simeq \int_{D_+} \frac{d^2 r}{a^2}\left[
\frac{\partial f(\nn_u)}{\partial \nn}\cdot(\be_v\times \nn_u) + \frac{\partial f(\nn_u)}{\partial \partial_i\nn}\cdot\partial_i(\be_v\times \nn_u)
-\frac{\partial f(\tt)}{\partial \nn}\cdot(\be_v^0\times \tt) - \frac{\partial f(\tt)}{\partial \partial_i\nn}\cdot\partial_i(\be_v^0\times \tt)
\right]\nonumber\\
&=\int_{D_+} \frac{d^2 r}{a^2}
\partial_i \left[ \frac{\partial f(\nn_u)}{\partial \partial_i\nn}\cdot(\be_v\times \nn_u)
 - \frac{\partial f(\tt)}{\partial \partial_i\nn}\cdot(\be_v^0\times \tt)
\right]\nonumber\\
&=\oint_{\partial D_+}\frac{d\ell_j}{a^2}\epsilon_{ij}\left[ \frac{\partial f(\nn_u)}{\partial \partial_i\nn}\cdot(\be_v\times \nn_u)
 - \frac{\partial f(\tt)}{\partial \partial_i\nn}\cdot(\be_v^0\times \tt)
\right],
\label{eq:V+approximation}
\end{align}
\end{widetext}
where $\epsilon_{ij}$ is the Levi-Civita symbol in two dimensions, $\partial D_+$ is the boundary of the area $D_+$, and $d\bm\ell$ is the vector element of line length.
Here, we have used the fact that $\nn_u(\rr)$ and $\tt$ satisfy Eq.~\eqref{eq:stationary_condition} from the third to the fourth lines
and the Green's theorem from the fourth to the fifth lines.

When the system is large enough, the integration along $\partial D_+$ vanishes except for the boundary between $D_+$ and $D_-$, since $\nn_u\to \tt$ and $\be_v,\be_v^0\to \bm 0$ as $\rr\to \infty$.
Thus, we obtain
\begin{align}
&V_+(\RR)\nonumber\\
&\simeq
\int_{-\Gamma}\frac{d\ell_j}{a^2}\epsilon_{ji}\left[ \frac{\partial f(\nn_u)}{\partial \partial_i\nn}\cdot(\be_v\times \nn_u)
 - \frac{\partial f(\tt)}{\partial \partial_i\nn}\cdot(\be_v^0\times \tt)
\right].
\label{eq:V+0}
\end{align}
When $|\RR|$ is large, 
we can further expand $\nn_u$ as $\nn_u\simeq\tt+\delta\nn_u$ on the boundary $\Gamma$, obtaining
\begin{align}
\frac{\partial f(\nn_u)}{\partial \partial_i n_\alpha} \simeq&
\frac{\partial f(\tt)}{\partial \partial_i n_\alpha} 
+\frac{\partial^2 f(\tt)}{\partial (\partial_i n_\alpha) \partial n_\beta} \delta n_{u,\beta}\nonumber\\
&+\frac{\partial^2 f(\tt)}{\partial (\partial_i n_\alpha) \partial(\partial_j n_\beta)} \partial_j \delta n_{u,\beta},\label{eq:expansion_f}\\
(\be_v\times \nn_u)_\alpha \simeq& \delta n_{v,\alpha} - (\delta\nn_u\cdot\delta \nn_v) \hat{t}_{\alpha},
\end{align}
where we have used Eq.~\eqref{eq:be_v}.
Substituting the above equations in Eq.~\eqref{eq:V+0}
we obtain
\begin{widetext}
\begin{align}
V_+(\RR)&\simeq
\int_{\Gamma}\frac{d\ell_j}{a^2}\epsilon_{ji}\left[
\frac{\partial f(\tt)}{\partial \partial_i n_\alpha} (\delta\nn_u\cdot\delta \nn_v) t_{\alpha}
-\frac{\partial^2 f(\tt)}{\partial (\partial_i n_\alpha) \partial n_\beta} \delta n_{u,\beta}\delta n_{v,\alpha}
-\frac{\partial^2 f(\tt)}{\partial (\partial_i n_\alpha) \partial(\partial_k n_\beta)} (\partial_k \delta n_{u,\beta}) \delta n_{v,\alpha}\right].
\label{eq:V+}
\end{align}
\end{widetext}

We note that 
the second-order terms of $\be_v$ and $\be_v^0$ neglected in the second line of Eq.~\eqref{eq:V+approximation} lead to contributions higher-order in $\delta\nn_{u,v}$ to $V_+(\RR)$, if $\nn_\textrm{1sk}$ rapidly converges to $\tt$ as in the case of the isotropic case under a vertical magnetic field, for which $\delta \nn_\textrm{1sk}(\rr)\propto e^{-\tilde{r}}/\sqrt{\tilde{r}}$ with $\tilde{r}=|\rr|/\sqrt{B/Ja^2}$.
To be more concrete, 
the next order terms to the second line of Eq.~\eqref{eq:V+approximation} are given by
\begin{widetext}
\begin{align}
    F_2=\int_{D_+}\frac{d^2r}{a^2} \bigg\{&
    \frac{\partial f(\nn_u)}{\partial \nn}\cdot \left[\frac{1}{2}\be_v\times (\be_v\times \nn_u)\right]+\frac{\partial f(\nn_u)}{\partial \partial_i\nn}\cdot \partial_i \left[\frac{1}{2}\be_v\times (\be_v\times \nn_u)\right]\nonumber\\
    &-\frac{\partial f(\tt)}{\partial \nn}\cdot \left[\frac{1}{2}\be_v^0\times (\be_v^0\times \tt)\right]-\frac{\partial f(\tt)}{\partial \partial_i\nn}\cdot \partial_i \left[\frac{1}{2}\be_v^0\times (\be_v^0\times \tt)\right]\nonumber\\
    &+\frac{\partial^2 f(\nn_u)}{\partial n_\alpha \partial n_\beta}(\be_v\times \nn_u)_\alpha(\be_v\times \nn_u)_\beta
    -\frac{\partial^2 f(\tt)}{\partial n_\alpha \partial n_\beta}(\be_v^0\times \tt)_\alpha(\be_v^0\times \tt)_\beta\nonumber\\
    &+2\frac{\partial^2 f(\nn_u)}{\partial \partial_i n_\alpha \partial  n_\beta} \left[\partial_i(\be_v\times \nn_u)_\alpha\right](\be_v\times \nn_u)_\beta
    -2\frac{\partial^2 f(\tt)}{\partial \partial_i n_\alpha \partial  n_\beta} \left[\partial_i(\be_v^0\times \tt)_\alpha\right](\be_v^0\times \tt)_\beta\nonumber\\
    &+\frac{\partial^2 f(\nn_u)}{\partial \partial_i n_\alpha \partial \partial_j n_\beta} \left[\partial_i(\be_v\times \nn_u)_\alpha\right]\left[\partial_j(\be_v\times \nn_u)_\beta\right]
    -\frac{\partial^2 f(\tt)}{\partial \partial_i n_\alpha \partial \partial_j n_\beta} \left[\partial_i(\be_v^0\times \tt)_\alpha\right]\left[\partial_j(\be_v^0\times \tt)_\beta\right]
    \bigg\}.
    \label{eq:F2}
\end{align}
\end{widetext}
Using Eq.~\eqref{eq:stationary_condition} and the Green's theorem, the first and second lines are rewritten as line integrals along $\Gamma$, which can be evaluated as in the above, resulting in the third order of $\delta\nn$.
On the other hand, We cannot rewrite the other three lines in simple line integrals.
However, in the case when $\be_v$ and $\be_v^0$ vanish as exponential functions of the distance $|\rr-\rr_v|$, 
the contribution to the area integral in $D_+$ mostly comes from the region close to the boundary $\Gamma$, where $\nn_u$ can be expanded as $\nn_u\simeq \tt+\delta\nn_u$.
We expand the derivatives of $f$ around $\nn_u\simeq \tt$ as in Eq.~\eqref{eq:expansion_f} and perform the subtraction in each of the last three lines, obtaining an additional factor $\delta\nn_u$. Thus, the contribution of $F_2$ in Eq.~\eqref{eq:F2} to $V_+(\RR)$ is in the third order of $\delta \nn_{u,v}$ and negligible to the leading terms given by Eq.~\eqref{eq:V+}.

Similarly, we calculate the approximate form for $V_-(\RR)$,
which is given by the right-hand side of Eq.~\eqref{eq:V+} with the replacements $\Gamma\to -\Gamma$ and $\delta\nn_u \leftrightarrow\delta\nn_v$.
As a whole, $V_\textrm{int}(\RR)$ in Eq.~\eqref{eq:Vint=V++V-} is approximated by
\begin{align}
    V_\textrm{app}(\RR)=&\int_{\Gamma}\frac{d\ell_j}{a^2}\epsilon_{ji}\left(A_{vu}-A_{uv}\right)_i,\\
    (A_{uv})_i=& \frac{\partial^2 f(\tt)}{\partial n_\alpha \partial (\partial_i n_\beta)} \delta n_{u,\alpha}\delta n_{v,\beta}\nonumber\\
&+\frac{\partial^2 f(\tt)}{\partial(\partial_k n_\alpha) \partial (\partial_i n_\beta)} (\partial_k \delta n_{u,\alpha}) \delta n_{v,\beta},
\end{align}
which are identical to Eqs.~\eqref{eq:Vint_J} and \eqref{eq:def_A}
under the replacements of $u\to +$ and $v\to -$.

\bibliography{reference}
\end{document}